%% file: main.tex
\documentclass[sigconf,authorversion,nonacm]{acmart}
\input{settings}

\AtBeginDocument{%
  }

\setcopyright{acmlicensed}
\copyrightyear{2018}
\acmYear{2018}
\acmDOI{XXXXXXX.XXXXXXX}
\acmConference[Conference acronym 'XX]{Make sure to enter the correct
  conference title from your rights confirmation email}{June 03--05,
  2018}{Woodstock, NY}
\acmISBN{978-1-4503-XXXX-X/2018/06}

\copyrightyear{2026}
\acmYear{2026}
\setcopyright{cc}
\setcctype{by}
\acmConference[CHI '26]{Proceedings of the 2026 CHI Conference on Human Factors in Computing Systems}{April 13--17, 2026}{Barcelona, Spain}
\acmBooktitle{Proceedings of the 2026 CHI Conference on Human Factors in Computing Systems (CHI '26), April 13--17, 2026, Barcelona, Spain}
\acmPrice{}
\acmDOI{10.1145/3772318.3790334}
\acmISBN{979-8-4007-2278-3/2026/04}

\begin{document}

\title{GeoVisA11y: An AI-based Geovisualization Question-Answering System for Screen-Reader Users}


\settopmatter{authorsperrow=4}

\author{Chu Li}
\orcid{0009-0003-7612-6224}
\affiliation{%
    \institution{Paul G. Allen School of Computer Science and Engineering}
  \institution{University of Washington}
  \city{Seattle}
\state{Washington}
  \country{USA}
}
\email{chuchuli@cs.washington.edu}

\author{Rock Yuren Pang}
\orcid{0000-0001-8613-498X}
\affiliation{%
    \institution{Paul G. Allen School of Computer Science and Engineering}
  \institution{University of Washington}
  \city{Seattle}
    \state{Washington}
  \country{USA}
}
\email{ypang2@cs.washington.edu}

\author{Arnavi Chheda-Kothary}
\orcid{0000-0001-8627-0412}
\affiliation{%
    \institution{Paul G. Allen School of Computer Science and Engineering}
  \institution{University of Washington}
  \city{Seattle}
  \state{Washington}
  \country{USA}
}
\email{chheda@cs.washington.edu}

\author{Ather Sharif}
\orcid{0000-0002-8729-1466}
\affiliation{%
    \institution{Paul G. Allen School of Computer Science and Engineering}
  \institution{University of Washington}
  \city{Seattle}
\state{Washington}
  \country{USA}
}
\email{asharif@cs.washington.edu}

\author{Henok Assalif}
\orcid{0009-0009-1227-4950}
\affiliation{%
    \institution{Paul G. Allen School of Computer Science and Engineering}
  \institution{University of Washington}
  \city{Seattle}
    \state{Washington}
  \country{USA}
}
\email{henok206@uw.edu}

\author{Jeffrey Heer}
\orcid{0000-0002-6175-1655}
\affiliation{%
    \institution{Paul G. Allen School of Computer Science and Engineering}
  \institution{University of Washington}
  \city{Seattle}
\state{Washington}
  \country{USA}
}
\email{jheer@cs.washington.edu}

\author{Jon E. Froehlich}
\orcid{0000-0001-8291-3353}
\affiliation{%
    \institution{Paul G. Allen School of Computer Science and Engineering}
  \institution{University of Washington}
  \city{Seattle}
  \state{Washington}
  \country{USA}
}
\email{jonf@cs.washington.edu}

\renewcommand{\shortauthors}{Li et al.}

\begin{abstract}
    \input{sections/0-abstract}
\end{abstract}



\begin{teaserfigure}
  \includegraphics[width=\textwidth]{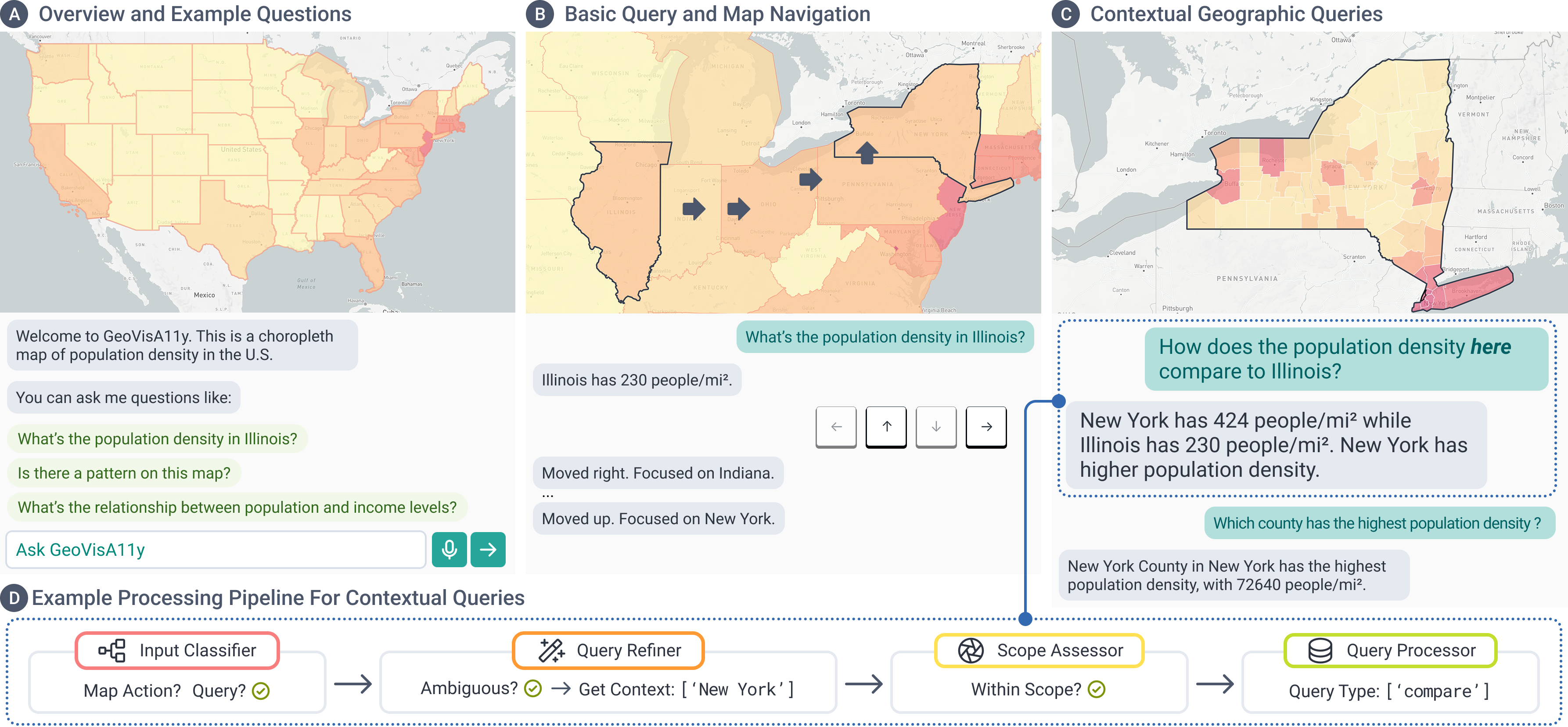}
  \caption{ 
    \sysname is an AI-based question-answering system for geovisualizations designed for screen-reader users.
    (A) On load, users see a visualization overview and example questions.
    (B) Location queries auto-focus the map; keyboard navigation also available.
    (C) Contextual queries answer based on current map focus, with automatic state-to-county level changes.
    (D) Pipeline showing how queries are classified, refined, assessed for scope, and processed to generate responses.
  }
  \Description{
  Four-panel diagram of GeoVisA11y, an AI question-answering system for accessible geovisualizations. Panel A shows the initial interface with visualization overview and example questions for users. Panel B demonstrates location querying that automatically focuses the map, with keyboard navigation controls. Panel C illustrates contextual queries with automatic geographic scale adjustment from state to county level. Panel D displays the system's processing pipeline, showing the flow from query classification through term refinement, scope assessment, to response generation.
  }
  \label{fig:teaser}
\end{teaserfigure}


\maketitle

\input{sections/1-intro}
\input{sections/2-related-work}

\input{sections/3-system-design}
\input{sections/4-system}
\input{sections/5-user-study}
\input{sections/6-findings}
\input{sections/7-discussion}
\input{sections/8-conclusion}
\input{sections/9-acks}

\bibliographystyle{ACM-Reference-Format}
\bibliography{references}

\end{document}

%% file: settings.tex
\PassOptionsToPackage{table,xcdraw}{xcolor} 
\usepackage{xcolor} 

\usepackage{dirtytalk} 
\usepackage[bottom]{footmisc} 
\usepackage{array, booktabs}
\usepackage{tabularray}
\usepackage{makecell}
\usepackage{graphicx}
\usepackage{multirow}
\usepackage{pifont}
\usepackage{colortbl}
\usepackage{stfloats}

\usepackage{fancybox}
\usepackage{xcolor}
\usepackage{textcomp}
\usepackage{amsmath}

\newcommand{\simplekey}[1]{\fbox{\texttt{\small #1}}}

\definecolor{myBlue}{RGB}{0, 122, 255}
\definecolor{myBlack}{RGB}{0, 0, 0}
\definecolor{myRed}{RGB}{190,55,74}

\newcommand{\rev}[1]{{\color{myBlack}{#1}}}

\newcommand{\sayit}[1]{{\textit{\say{#1}}}}

\newcommand{\sayhi}[1]{{\textit{\say{#1}}}}

\usepackage{xspace}

\def\sysname{GeoVisA11y\xspace}

%% file: sections/0-abstract.tex
Geovisualizations are powerful tools for communicating spatial information, but are inaccessible to screen-reader users. 
To address this limitation, we present \sysname, an LLM-based question-answering system that makes geovisualizations accessible through natural language interaction. 
The system supports map reading, analysis, interpretation and navigation by handling analytical, geospatial, visual, and contextual queries. 
Through user studies with \rev{six} screen-reader users and \rev{six}  sighted participants, we demonstrate that \sysname effectively bridges accessibility gaps while revealing distinct interaction patterns between user groups. We contribute: (1) an open-source, accessible geovisualization system, (2) empirical findings on query and navigation differences, and (3) a dataset of geospatial queries to inform future research on accessible data visualization.

%% file: sections/1-intro.tex
\section{Introduction}
\begin{quote}
  \sayhi{[GeoVisA11y] made it all come to life. Instead of having to muddle through a bunch of numbers and try to find correlations myself, the way it could analyze information and describe patterns is impressive.} 
  \newline
  --B5, a blind participant
\end{quote}

Interactive map visualizations, or \textit{geovisualizations}, are powerful tools for discerning patterns, trends, and relationships in spatial data~\cite{coltekin_geovisualization_2018, andrienko_exploratory_2006}.
Geovisualizations support informed decision-making across urban planning~\cite{yeh_urban_1999} and public health~\cite{boscoe_choosing_2003}. 
They also communicate essential information in news media, such as COVID case numbers~\cite{times_coronavirus_2020} and election results~\cite{park_extremely_2021, datar_extremely_2025}.  
Yet, most geovisualizations are inaccessible to screen-reader users~\cite{sharif_understanding_2022, zong_rich_2022, fan_accessibility_2023}.
Even when accessibility features such as \textit{alt text} and \textit{data tables} are available, they fail to capture the full analytical and interpretive potential of geovisualizations~\cite{gorniak_vizability_2024, kim_beyond_2023}. 
A study of fifteen geovisualizations found that \textit{none} communicated any high-level spatial patterns to screen-reader users~\cite{fan_accessibility_2023}.

While recent work like \textit{AltGeoViz}~\cite{li_altgeoviz_2024} and \textit{Data Navigator}~\cite{elavsky_data_2023}, offer alt text for geovisualizations based on user interactions, these approaches remain largely descriptive, akin to ``map reading'' rather than ``map analysis''~\cite{mersy_choropleth_1990, muehrcke_functional_1978, olson_coordinated_1976}. 
We aim to support deeper spatial interpretation---allowing screen-reader users to identify patterns, relationships, and geometric characteristics that are crucial for insight building~\cite{mersy_choropleth_1990, muehrcke_functional_1978, olson_coordinated_1976}.
Emerging question-answering (QA) systems for geovisualizations such as \textit{VoxLens}~\cite{sharif_voxlens_2022} represent progress but rely on keyword matching, limiting interactive queries. 

In this paper, we introduce \sysname, an AI-based geovisualization QA system that advances the field by enabling complex spatial interpretation and analysis for screen-reader users. 
By integrating geostatistical analysis with large language models, \sysname moves beyond keyword-matching and rule-based approaches to support previously inaccessible analytical tasks.
Moreover, \sysname uniquely handles geospatial queries about \textit{spatial patterns}~\cite{schiewe_empirical_2019}, \textit{spatial relationships}, and \textit{geometric characteristics}.
To ease interaction, users can fluidly switch between keyboard-based navigation and conversational commands for map exploration. 
\sysname was designed iteratively in collaboration with two screen-reader users and drew upon design principles from accessible QA literature, including disambiguating deictic references~\cite{kim_exploring_2023, hoque_applying_2018} and supporting general/contextual queries~\cite{kim_exploring_2023, gorniak_vizability_2024}.

Compared with simple data visualizations such as bar charts, line graphs and scatter plots, geovisualizations are inherently more complex: scale, distance, location and direction all carry complex real-world meanings~\cite{maceachren_research_2000, andrienko_geovisual_2007, anselin_what_1989, fan_understanding_2024}. 
This visual and semantic complexity affects all users: even sighted individuals misinterpret geovisualizations due to misrepresented scales, distorted projections, misleading color schemes, and inappropriate classification methods~\cite{schiewe_empirical_2019, juergens_trustworthy_2020, monmonier_how_2018}.
Following the universal design paradigm~\cite{steinfeld_universal_2012}, where accessibility features initially developed for specific populations subsequently benefit broader populations (\textit{e.g.,} subtitles for deaf and hard-of-hearing users, curb ramps for wheelchair users), we hypothesize that accessible geospatial interfaces will serve both screen-reader users and sighted users experiencing difficulty with complex geovisualizations. 
Building on this motivation, we use \sysname to investigate the following research questions:
\begin{itemize}
    \item [\textbf{RQ1}] What \textit{types of queries} do screen-reader users make when interacting with a geovisualization QA system?
    \item [\textbf{RQ2}] How can a geovisualization QA system effectively support \textit{screen-reader users} in engaging with geovisualizations?
    \item [\textbf{RQ3}] To what extent can such a system enhance \textit{sighted users' }engagement with geovisualizations?
    \item [\textbf{RQ4}] What are the key differences in \textit{querying strategies and interaction patterns} between sighted and screen-reader users when using such a system?
\end{itemize}

To answer our research questions, we conducted a user study with six screen-reader users and six sighted participants. 
Importantly, our goal is not to position the sighted users as a \textit{baseline} group~\cite{mack_what_2021} to derive confirmatory conclusions, but rather to explore how both groups interact with \sysname and investigate how visualization QA tools designed for screen-reader users could benefit broader populations.
Participants performed exploratory data analysis for two tasks: (1) distributing digital equity funding based on underserved populations and limited digital access, and (2) identifying predominant energy sources across U.S. regions and their potential causes.
We found that \sysname effectively supports both user groups across the three levels of geovisualization engagement (map \textit{reading, analysis, interpretation}~\cite{muehrcke_functional_1978}).
Screen-reader participants successfully employed diverse querying strategies for spatial analysis and valued \sysname's informative and clear responses. 
They particularly highlighted how the system enabled self-directed spatial exploration.
Sighted users demonstrated varying levels of system usage based on their domain expertise and geographic knowledge, with increased usage when working with more complex visualizations.
For querying strategies, screen-reader users relied primarily on verbal queries combined with keyboard navigation, whereas sighted users typically performed visual assessment before querying specific details.
Despite the different approaches, participants from both groups successfully identified similar patterns in the data, demonstrating that \sysname represents a step toward creating a shared understanding of geovisualizations.
We also observed differences in how information was processed: screen-reader users tended to place greater trust in system-provided descriptions, while sighted users' judgments appeared influenced by visual biases such as region size.
These findings reveal opportunities to further bridge perceptual gaps and create more robust shared understanding of geovisualizations.

In summary, we contribute: 
(1) \textit{\sysname}, an open-source system for accessible geovisualization interaction\footnote{Our repository is available at \url{https://github.com/makeabilitylab/geovisa11y}.};
(2) \textit{an empirical qualitative evaluation} with 12 participants that highlights the differences between the blind and low-vision (BLV)\footnote{In this paper, we use \textit{`BLV'} and \textit{`screen-reader user'} interchangeably.} and sighted user groups in terms of query formation and map navigation; and
(3) a \textit{dataset} containing queries asked by both user cohorts to inform future research in accessible, geovisualization QA tools.

%% file: sections/2-related-work.tex
\section{Related Work}
Our work builds on prior work in geospatial visualization, accessible data visualization, and visualization QA systems.

\subsection{Geospatial Visualization and Analysis}
MacEachren \textit{et al.}~\cite{maceachren_visualization_1992} defined geovisualization as using concrete visual representations to make spatial contexts and problems visible.
Scholars have established three progressive levels of engagement:
(1) \textit{map reading}: identifying what elements the cartographer has encoded, 
(2) \textit{map analysis}: recognizing significant patterns among these elements, and
(3) \textit{map interpretation}: explaining observed patterns by integrating additional knowledge~\cite{mersy_choropleth_1990, muehrcke_functional_1978, olson_coordinated_1976}.
The engagement framework highlights pattern recognition as a critical step from basic reading to comprehensive analysis.
However, the concept of a \textit{pattern} remains challenging to define: it is often subjectively understood and intuitively recognized, with scholars frequently discussing patterns without explicit definitions (\textit{e.g.}~\cite{slocum_thematic_2022, mersy_choropleth_1990}).

While verbalizing visual patterns is essential for screen-reader users, explicit pattern descriptions benefit all users. 
Geovisualizations are more complex than traditional visualizations, making them more susceptible to misinterpretation~\cite{slocum_thematic_2022, monmonier_how_2018} and requiring geospatial data literacy for meaningful interaction~\cite{juergens_digital_2020, schiewe_empirical_2019}.
\citet{schiewe_empirical_2019} identified three perceptual biases in choropleth map interpretation: associating darker colors with higher values, neglecting of smaller geographic areas, and misinterpreting patterns based on data classification methods.
Despite the need for pattern articulation, few studies have generated comprehensive textual descriptions that fully capture visual patterns~\cite{li_altgeoviz_2024,thomas_atlastxt_2007, thomas_whats_2008}. 
AltGeoViz~\cite{li_altgeoviz_2024} generates dynamic alt text of basic statistics (min, max, average), viewport boundaries, zoom levels, and spatial patterns using grid-based techniques. 
However, it does not account for polygon geometry or spatial autocorrelation.
Atlas.txt~\cite{thomas_whats_2008, thomas_atlastxt_2007} offers comprehensive descriptions including maximum/minimum clusters, trends, averages, boundary information, human/physical geography, and cardinal directions. 
Despite these capabilities, their clustering and trend identification methods lack documentation, and the system does not support interactive exploration or query-based interactions.
\sysname addresses this gap by adopting Schiewe's~\cite{schiewe_empirical_2019} definition of \textit{pattern} as the \textit{detection of hot/cold spots}, and implementing Local Indicators of Spatial Association (LISA)~\cite{anselin_local_1995}, a widely used geostatistical approach to identify geographical clustering of high and low values.

\subsection{Accessible Data Visualization}
Research on visualization accessibility has introduced various modalities, including tactile, haptic and multimodal techniques~\cite{kim_accessible_2021}, but these custom hardware solutions remain difficult to access.
Screen-readers continue to be the most widely adopted and affordable assistive technology~\cite{kim_accessible_2021}.
For online visualizations, data tables and alt text are the most commonly used and recommended accommodations for screen-reader users~\cite{zong_rich_2022,wgbh_national_center_for_accessible_media_effective_2025, initiative_wai_complex_2022, choi_visualizing_2019}. 
However, both approaches have significant limitations. 
Data tables lose the abstraction benefits of visualizations, making it difficult to identify broader patterns or trends~\cite{zong_rich_2022}.
Static alt text requires users to accept the creator's interpretation without allowing personal exploration~\cite{lundgard_accessible_2022}.
More recent approaches include automatic alt text generation through computational pipelines that extract visualization features~\cite{mishra_chartvi_2022, choi_visualizing_2019}. 
Despite these advances, alt-text-based solutions remain largely unidirectional, \textit{i.e., }systems present information to users who must passively accept it. 

Question-answering (QA) capabilities overcome this unidirectional limitation by letting users actively query the system.
Sharif \textit{et al.} developed \textit{VoxLens}~\cite{sharif_voxlens_2022}, a JavaScript plugin whose QA module responds to predefined keywords like "maximum," "minimum," and "median." In subsequent work, they extended VoxLens to support geospatial visualization queries and more detailed information extraction~\cite{sharif_understanding_2022,sharif_understanding_2023}.
These approaches were limited by the technology available at the time, supporting only simple, keyword-based queries rather than flexible, robust interactions.
Recent AI advances and improved model accessibility enable more responsive systems.
Our work builds on this foundation by making geovisualizations accessible to BLV users through natural language interactions.

\begin{figure*}[t]
    \centering
    \includegraphics[width=\linewidth]{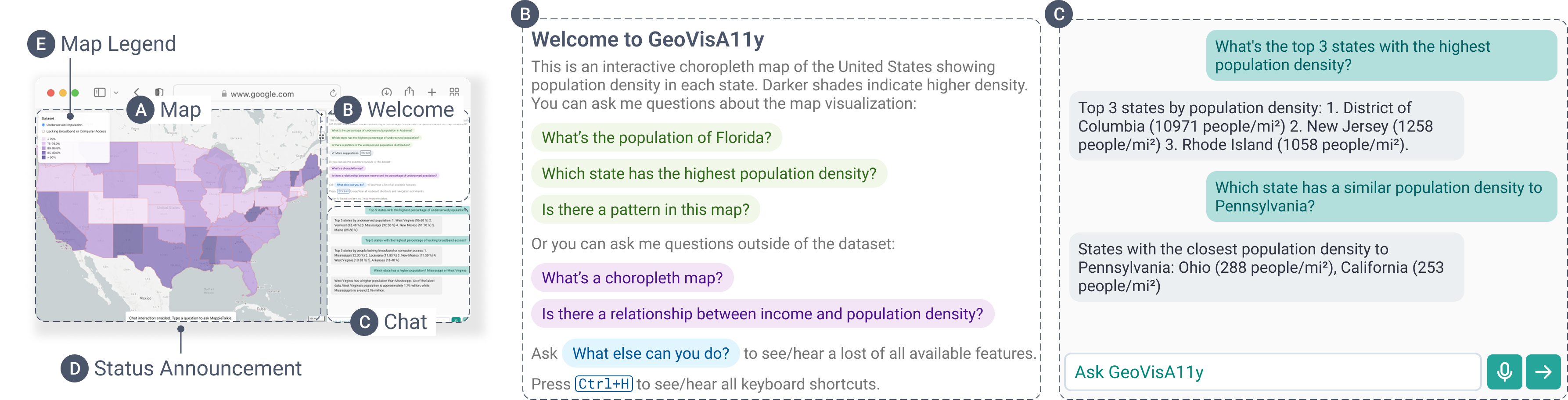}
    \caption{
    \sysname UI overview. (A) Interactive map component. (B) Chat component where users can ask questions and receive answers through text or voice input. (C) Welcome section with an overview of the visualization, example questions, and shortcut key instructions. (D) Status indicator that announces changes to screen-reader users, \textit{e.g., ``Map interaction enabled. Focused on Kansas.''} or \textit{``Chat interaction enabled. Type a question to ask \sysname.''} (E) Legend of the current interactive map.
    }
    \Description{
    User interface overview showing five main components: (A) Interactive map component in the center, (B) Chat component allowing users to ask questions via text or voice input, (C) Welcome section displaying visualization overview, example questions, and keyboard shortcut instructions, (D) Status indicator that provides audio announcements to screen-reader users such as 'Map interaction enabled. Focused on Kansas' or 'Chat interaction enabled. Type a question to ask the system', and (E) Legend explaining the current interactive map symbols and data.
    }
    \label{fig:ui}
\end{figure*}

\subsection{Visualization Question-Answering} 
QA systems for data visualization have gained popularity in helping users understand and complete complex analytical tasks~\cite{kim_answering_2020, setlur_eviza_2016, hoque_applying_2018, gao_datatone_2015,sharif_voxlens_2022}.
These systems respond to natural language queries by offering direct answers~\cite{sharif_voxlens_2022, kim_answering_2020}, highlighting relevant information in existing visualizations~\cite{setlur_eviza_2016, hoque_applying_2018}, or creating new visualizations tailored to the query~\cite{gao_datatone_2015}.
For geovisualizations specifically, \textit{MapQA} offers a large-scale dataset of approximately 800K QA pairs covering about 60K choropleth maps~\cite{chang_mapqa_2022}.
Despite the scale, MapQA addresses only three basic question categories—legend interpretation, data retrieval, and relational queries—none going beyond the elementary \textit{map reading} stage of geovisualization engagement.
Furthermore, these systems depend on
vision-language models, which are more susceptible to errors as visualization QA is particularly vulnerable to slight pixel-level variations compared to natural image QA~\cite{bursztyn_representing_2024}.
Recent studies demonstrate that incorporating chart specifications directly into language models yields superior performance compared to vision-language approaches~\cite{bursztyn_representing_2024,gorniak_vizability_2024}.
Hence, \sysname employs a language-based pipeline to achieve greater accuracy and reliability.
LLM-based systems are particularly promising for geovisualization QA because they inherently possess geographic knowledge that enables understanding of how geospatial insights emerge from the interplay of multiple contextual layers–such as population density, infrastructure, and natural features~\cite{mai_opportunities_2024, lietard_language_2021}.
Moreover, while vision-language models struggle with spatial reasoning, LLMs paired with structured geographic data can achieve precise spatial analysis by directly processing coordinate information, topological relationships, and spatial autocorrelation measures rather than relying on less reliable visual inference.

Although visualization QA systems often cite accessibility as a potential application, they rarely focus on BLV users' specific needs, incorrectly assuming their questioning behavior mirrors that of sighted users~\cite{kim_exploring_2023}.
Research in both image and visualization QA has rejected this assumption, demonstrating that BLV users pose significantly different questions with distinct phrasing patterns~\cite{dognin_image_2022,gurari_vizwiz_2018,kim_exploring_2023}.
To address this gap, Kim \textit{et al.}~\cite{kim_exploring_2023} conducted a Wizard of Oz study with BLV participants and developed design considerations for accessible visualization QA systems. 
\textit{VizAbility}~\cite{gorniak_vizability_2024} represents perhaps the closest approach to our work—a QA system that enhances chart content navigation through conversational interaction, allowing users to query visual data trends using natural language. 
While VizAbility can process choropleth maps, it works only with static visualizations; our work differs by focusing on interactive, comprehensive geovisualizations.

%% file: sections/3-system-design.tex
\section{Designing \sysname}
\begin{figure*}[t]
    \centering
    \includegraphics[width=\linewidth]{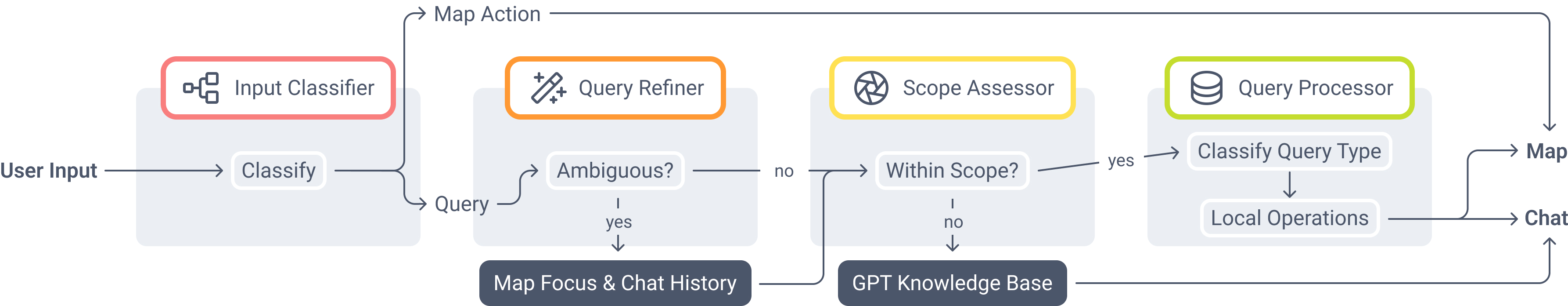}
    \caption{
    System pipeline diagram. 
    All four components use GPT-4o-mini via few-shot prompting.
    (1) The Input Classifier separates map action commands from information queries.
    (2) The Query Refiner resolves ambiguities (\textit{e.g.,} replacing \textit{``here''} with \textit{``Ohio''}, \textit{``it''} with \textit{``population density''}) by ingesting current map focus and chat history. 
    (3) The Scope Assessor determines if the query can be answered from the local database.
    (4) The Query Processor classifies the query and triggers relevant operations, then updates the Chat and/or Map interface.
    }
    \Description{
    System pipeline flowchart showing user input processing through four main components. Flow begins with user input going to Input Classifier (red box), then to Query Refiner (orange box) which checks if query is ambiguous - if yes, adds context; if no, proceeds to Scope Assessor (yellow box) which determines if query is within scope - if no, routes to GPT-4o-mini; if yes, continues to Query Processor (green box) that classifies query type and performs local operations, finally outputting to either Map or Chat interface.
    }
    \label{fig:pipeline}
\end{figure*}
To design and build \sysname, we drew on the above literature and followed an iterative, human centered design process involving six hours of co-design sessions with two blind participants over three weeks. 
Our early-stage interactive prototype consisted of a chat interface with basic AI query capabilities and an interactive map that responded to chat-based navigation commands.
We selected participants who exemplify our target user population: individuals who use screen-readers to actively engage with data visualization and analysis in professional or personal contexts.
Participants included one individual blind from birth (C1) and another who experienced gradual vision loss and is now completely blind (C2), representing different experiences with vision loss (\autoref{tab:participant}).  
Participants  were compensated \$40 per hour.

\subsection{System Iteration}
Through our iterative co-design process, we identified key areas for system improvement across query handling, navigation paradigms, and interface design.

\textbf{Initial Reactions.} 
Both co-design participants had prior experience with online map visualizations but found them challenging to access. 
C1 noted: \sayhi{I really liked how intuitive \sysname was. It surprised me... I threw some more nuanced questions at it and it understood.}
When asked about the usefulness of the prototype, C1 explained: \sayhi{I think it could definitely improve my geography knowledge, and just being able to read and understand [map visualizations]. Because before, reading maps was always a big undertaking.}
C2 stated \sayhi{I love this} and \sayhi{I could see so much potential in this, for people who are visually impaired like myself. Even people who are sighted can use this because it's a great way to access data points easily without having to sift through the traditional statistical maps.}

\textbf{Query Types.} 
While our initial prototype could handle query types identified by prior work (retrieve, compare, find extremum, aggregate, filter, sort, cluster)~\cite{kim_exploring_2023} and summarize spatial patterns, our co-design sessions revealed additional geovisualization-specific query types. 
These included queries about \textit{shapes} of geographic entities (\textit{e.g., \sayit{What is the shape of Alabama}}) and \textit{spatial relationships} between geographic entities (\textit{e.g., \sayit{What are the neighboring states of Tennessee?}}).
Participants also used \textit{context-aware querying}, asking questions like \sayit{What's the population density here?} without explicitly specifying the geographic context when they had already focused on a particular region.

\textbf{Map Navigation.} 
We began our co-design sessions with a map display that updated based on queries but did not support direct user interaction (\textit{e.g.,} pan, zoom, select).
C1 explicitly requested \textit{simple keyboard-based navigation mechanisms} to explore different states and counties. 
Prior research~\cite{li_altgeoviz_2024} shows screen-reader users expect arrow keys to facilitate discrete movement between regions (\textit{e.g.,} jumping from one state to its neighbor) rather than continuously panning as in interfaces designed for sighted visual exploration, so we implemented this discrete navigation approach using arrow keys.
Participants also demonstrated preferences for natural language navigation, typing commands like \sayhi{Focus on Washington} or \sayhi{Take me to Massachusetts} and expecting the map to update accordingly.

\subsection{Design Considerations}
From our co-design sessions and key literature, we synthesized the following design considerations:
\newline
\textbf{Support geovisualization-specific questions.}
Beyond query types identified by prior research~\cite{kim_exploring_2023,gorniak_vizability_2024}, our co-design revealed the need for queries about geographic entity shapes and spatial relationships between regions. 
In addition, to create a robust visualization QA experience for screen-reader users, an accessible geoanalytical system should handle disambiguation, context tracking, guided questions, semantic breadth, and contextual querying.
\newline
\textbf{Support direct and accessible map exploration.}
Navigation should align with screen-reader users' mental models, enabling discrete movement between geographic regions rather than continuous panning designed for visual exploration.
\newline
\textbf{Support synchronization between map and chat.}
The map and chat components should work in concert through synchronous focus (map updates based on chat interactions), natural language-driven navigation, and context-aware querying that supports implicit references to focused geographic entities.
\newline
\textbf{Follow standard accessibility guidelines.}
 In line with WCAG~\cite{w3c_web_2024} principles, designs should prioritize clear focus state awareness, consistent interaction feedback, flexible input modalities, accessible conversation history, plain language communication, and intuitive keyboard navigation patterns.

%% file: sections/4-system.tex
\section{The \sysname System}
\input{tables/table-query-types}
Informed by our iterative co-design process and prior work, we designed \sysname to make geovisual analytics accessible to screen-reader users through natural language interaction. 
\sysname has two primary components:
(1) a screen-reader compatible UI with an interactive map and an AI-based chat that supports analytical, geospatial, visual, and contextual queries; 
(2) a custom QA pipeline that combines chat questions with map interactions to form queries and uniquely combines geo-statistical analysis with LLM-based summaries.
We begin by describing the QA pipeline as it is a central technical contribution of our work.
\sysname is available as open-source software at: \url{https://github.com/makeabilitylab/geovisa11y}.

\subsection{QA Pipeline}
To transform natural language questions into geoanalytical responses and map interactions, we implemented a structured pipeline that parses user queries, determines appropriate response types, and coordinates both analytical processing and map state updates.
We first detail the pipeline's four-component architecture, then examine how each query type is processed and analyzed.

\subsubsection{Pipeline Architecture}

The \sysname QA pipeline consists of four components (\autoref{fig:pipeline}): the \textit{Input Classifier, Query Refiner, Scope Assessor, and Query Processor}.
All components use GPT-4o-mini with few-shot learning~\cite{brown_language_2020}, which offers strong classification performance~\cite{gao_making_2021, brown_language_2020}, low computational cost for rapid prototyping, and inference latency suitable for real-time applications~\cite{openai_gpt-4o_2024}.
We acknowledge that fine-tuning these models (\textit{e.g.,} on our dataset artifact) may improve task performance, but we leave this exploration to future work.
Additionally, we designed separate prompts for each component as LLM performance degrades with longer input length~\cite{levy_same_2024, liu_lost_2024}.
See Supplementary Materials for specific prompts.

\textbf{Input Classifier.} 
Upon receiving text or voice input from users, the Input Classifier employs few-shot prompting to categorize the input as either an \textit{action command} or an \textit{information query.} 
Action commands trigger direct map manipulation, information queries proceed to Query Refiner for further processing.

\textbf{Query Refiner.} 
This component addresses potential ambiguities in user queries. 
The system identifies and resolves two types of ambiguity: 
(1) \textit{location ambiguity}, where deictic references such as \sayit{here} or \sayit{this state} are resolved using the current map focus or previous conversation, \textit{e.g.,} \sayit{What's the population density here?} when a specific state is highlighted; 
and (2) \textit{topic ambiguity}, where pronouns like \sayit{that} or \sayit{it} are resolved using conversation history, \textit{e.g.,} \sayit{How does that compare to Ohio?} where \sayit{that} refers to previously discussed population density.  
Some queries exhibit both types, \textit{e.g.,} \sayit{How does it compare to its neighbors?} where \sayit{it} refers to a previously discussed metric and \sayit{its} refers to the focused state. 

\textbf{Scope Assessor.} 
Following disambiguation, the system evaluates whether the query falls within the scope of local data capabilities.
The system provides GPT with a compact textual description of the dataset schema (\textit{e.g.,} variable names, metric definitions, and geographic units).
Using this information, GPT acts as a binary classifier to determine whether the query can be answered using local data and geostatistical operations.
\textit{Within-scope queries} proceed to the Query Processor for classification and local operations, while \textit{beyond-scope queries} are routed to GPT's knowledge base.

\textbf{Query Processor.} 
The Query Processor uses GPT with few-shot prompting to classify queries into one of 14 predefined categories (\autoref{tab:query-types}). 
As with the Scope Assessor, no map imagery or raw datasets are sent to GPT—it serves strictly as a text-based classifier. 
Once classified, \textit{analytical} and \textit{geospatial} queries are processed using local database operations and spatial analysis algorithms.
This design ensures consistent, reproducible results with low latency. 
Only \textit{visual} and \textit{contextual} queries that cannot be answered by the local database are handled by GPT's knowledge base.
Section \ref{sec:query-taxonomy} details the processing approaches for each query type.

\textbf{Output Generation.} 
The pipeline delivers final results through coordinated updates to both interface components. 
Textual answers appear in the chat interface, and for locally processed queries, the system also highlights relevant geographic entities on the map.

\subsubsection{Query Taxonomy \& Analysis Approach}
\label{sec:query-taxonomy}

Building upon prior work in accessible chart QA system~\cite{kim_exploring_2023}, we identified five primary query categories for geovisualization interactions. 
See \autoref{tab:query-types} for supported query types and example questions.
Here, we detail the data analysis approach for each query type.

\textbf{Map Actions} directly manipulate the map visualization, letting users navigate to specific geographic locations (\textit{e.g.,} \sayit{Go to Ohio}).
The system processes these commands by looking up the target location's coordinates and repositioning the map view accordingly.
Screen-reader users receive audio announcements of their current location (\textit{e.g.,} \sayit{Now focused on Ohio.})

\textbf{Analytical Queries} include: (1) \texttt{retrieval} queries that access specific data points, (2) \texttt{comparison} queries that evaluate relationships between entities, (3) \texttt{extremum} queries that identify maximum or minimum values, (4) \texttt{aggregation} queries that compute summary statistics, (5) \texttt{filter} queries that identify geographic entities meeting specific criteria, (6) \texttt{sort} queries that rank geographic entities, and (7) \texttt{cluster} queries that identify similar value entities. 

Most analytical queries are processed through direct database operations, with results formatted using  templates that include metric units and contextual information.
Filter and sort queries first use GPT to extract numerical conditions or result counts (\textit{e.g.,} extracting \sayhi{>100} from \sayhi{Which states have density greater than 100?}) before executing database operations.
For cluster queries, the system identifies entities with values within 20\% margin of a reference entity.

\textbf{Geospatial Queries} include: (1) \texttt{pattern} identification queries that describe overall geographic trends and (2) \texttt{outlier} identification queries that highlight anomalous regions.

To summarize visual spatial patterns, we integrated global pattern assessment with local anomaly detection. Global pattern identification uses \textit{Moran's I}~\cite{moran_interpretation_1948} analysis to assess spatial autocorrelation.
Moran's I is a statistical measure that quantifies the degree of spatial clustering, indicating whether similar values tend to be located near each other more than expected by chance~\cite{moran_interpretation_1948}.
Our pipeline constructs spatial weight matrices using Queen contiguity~\cite{anselin_spatial_1988}, then computes Moran's I with 999 permutations to determine statistical significance. 
Results are interpreted as clustered ($I > 0, p <. 05$), dispersed ($I < 0, p < .05$), or random patterns ($p >. 05$).

\begin{figure*}[]
    \centering
    \includegraphics[width=\linewidth]{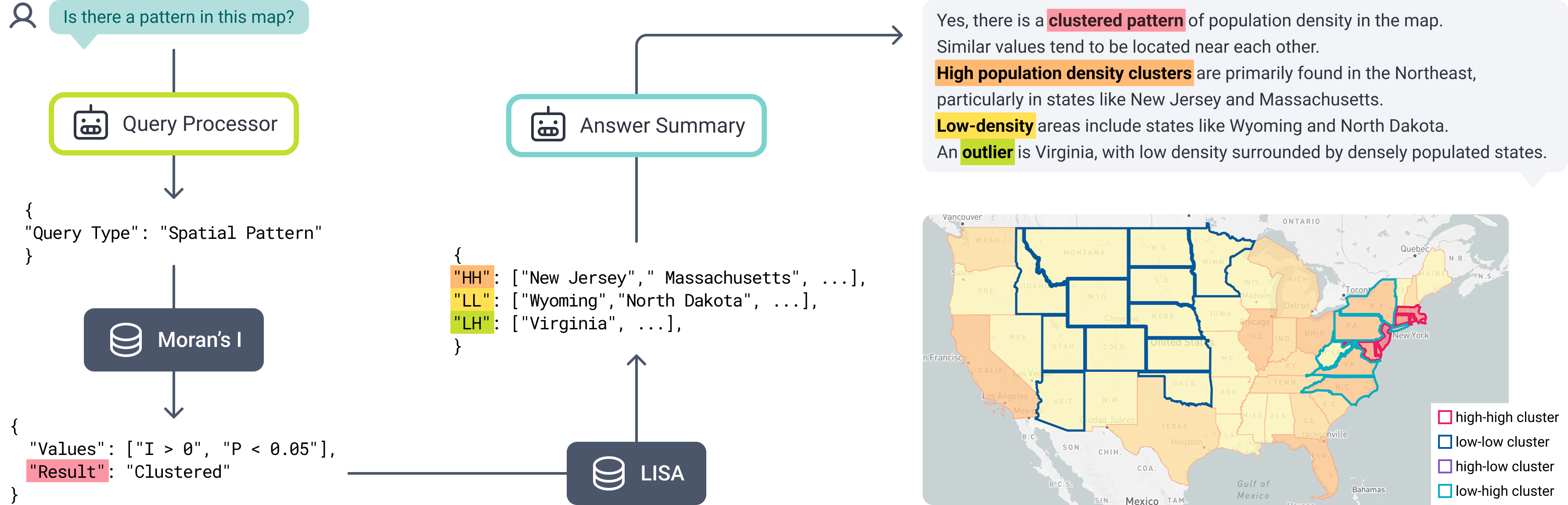}
    \caption{
    Pipeline workflow for spatial pattern queries. When users ask \textit{``Is there a pattern on the map?''}, the system runs global Moran's I analysis to detect significant patterns, then performs LISA cluster analysis if patterns exist. 
    LISA outputs are summarized by GPT with representative examples, while the map displays colored outlines for cluster types.
    \rev{
    Color highlights in the answer text indicate which part of the pipeline produced that information: the red highlight on ``clustered pattern'' corresponds to the Moran's I result, while orange, yellow, and green highlights correspond to specific cluster types (HH, LL, LH) identified by the LISA analysis.
    }
    }
    \Description{
    Pipeline workflow diagram showing how the system processes spatial pattern queries. User asks 'Is there a pattern in this map?' which flows through Query Processor (classified as 'Spatial Pattern'), then to Moran's I statistical analysis showing clustered results (I > 0, P < 0.05), followed by LISA cluster analysis identifying High-High clusters (New Jersey, Massachusetts), Low-Low clusters (Wyoming, North Dakota), and Low-High outliers (Virginia). The workflow generates both an Answer Summary explaining the clustered population density pattern and an updated map of the United States with colored regions indicating four cluster types: high-high (pink), low-low (blue), high-low (yellow), and low-high (teal) clusters.
    }
    \label{fig:pattern}
\end{figure*}

Outlier identification employs \textit{Local Indicators of Spatial Association} (LISA)~\cite{anselin_local_1995} analysis, an exploratory tool used to statistically assess geographic clustering of high and low values in a dataset. 
LISA calculates local spatial autocorrelation at each individual location using a single variable, enabling the quantitative estimation of local spatial clustering by indicating how similar an observation is to all other observations within a defined region~\cite{anselin_local_1995}.
LISA identifies four cluster types: \textit{High-High} (high values surrounded by high values), \textit{Low-Low} (low values surrounded by low values), \textit{High-Low} (high values surrounded by low values), and \textit{Low-High} (low values surrounded by high values)~\cite{anselin_local_1995}.
Only statistically significant clusters ($p < .05$) are reported. 
\sysname then synthesizes these statistical results into natural language summaries, providing two representative examples from each cluster type when available.
The map synchronously displays outlines highlighting the four cluster types.
This pipeline is illustrated in \autoref{fig:pattern}.

\textbf{Visual Queries} address the visual elements of the geovisualization rather than the underlying data, including: (1) \textit{legend or color queries}, 
(2) \textit{shapes of geographic entities queries},
and (3) \textit{spatial relationship queries} that explore the layout and adjacency of geographic entities. 
Legend queries are processed by providing GPT with information about legend color mappings and value specifications from the current map visualization. 
Shape and spatial relationship queries are handled entirely by GPT using its built-in geographic knowledge, as these queries typically concern general geographic facts rather than dataset-specific information.

\textbf{Contextual queries} are questions that extend beyond the immediate dataset, including: 
(1) \textit{visualization concept queries}, and (2) \textit{general knowledge queries} that connect the visualization to broader concepts.
All contextual queries are processed by GPT, prompted to act as a geovisualization expert providing concise answers that connect the visualization to broader geographic knowledge and visualization principles.

\input{tables/table-shortcut}

\subsection{User Interface}
\sysname's UI comprises a map visualization and an AI chat subsystem---both feed into the QA pipeline. 
Screen readers parse the map and messages and convey them through audio to participants.
Though customizable for other regions, we use a sample dataset about U.S. population density to demonstrate the system's functionality in this section. 
See~\autoref{fig:ui} and video for demonstration, and~\autoref{tab:shortcut} for all available shortcuts and functions.

\begin{figure*}[t]
    \centering
    \includegraphics[width=\linewidth]{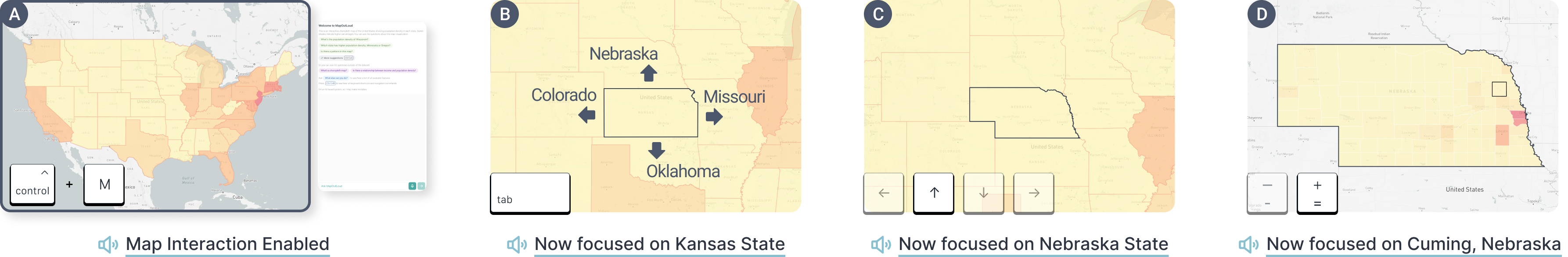}
    \caption{
    Interactive map keyboard navigation.
    (A) Ctrl + M activates the map interface.
    (B) TAB focuses on an initial state, then arrow keys navigate to adjacent states in cardinal directions.
    (C) Example navigation from Kansas to Nebraska using the up arrow.
    (D) + and - keys zoom between county and state levels, shown here focused on Cuming County, Nebraska.
    }
    \Description{
    Four-panel demonstration of interactive map keyboard navigation. Panel A shows the full United States map with keyboard controls visible (Ctrl and M keys) and status message 'Map Interaction Enabled.' Panel B displays a regional view of central states with Kansas highlighted and status 'Now focused on Kansas State.' Panel C shows the same regional view with focus shifted to Nebraska and status 'Now focused on Nebraska State.' Panel D shows a zoomed-in county-level view of Nebraska with status 'Now focused on Cuming, Nebraska.' Navigation controls show Ctrl+M activates the map, TAB focuses on initial state, arrow keys move between adjacent states, and +/- keys zoom between state and county levels.
    }
    \label{fig:map-interactions}
\end{figure*}

\begin{figure*}[b]
    \centering
    \includegraphics[width=\linewidth]{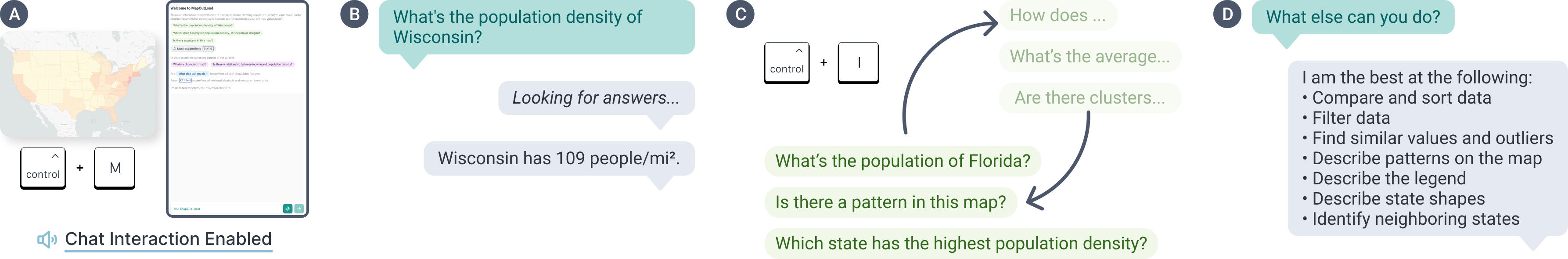}
    \caption{
    AI chat key interactions. (A) Ctrl + M activates the chat if previously focused on the map. 
    (B) Upon receiving a query, system repeats the user query followed by a status indicator.
    (C) The three example questions can be refreshed through Ctrl + I.
    (D) Asking \textit{``What else can you do?''} can see/hear all the available query types.
    }
    \label{fig:chat-interactions}
    \Description{
    Four-panel demonstration of AI chat interactions. Panel A shows the interface with status 'Chat Interaction Enabled' and Ctrl+M shortcut to activate chat from map view. Panel B displays a user query 'What's the population density of Wisconsin?' with system response showing 'Looking for answers...' status followed by the answer 'Wisconsin has 109 people/mi².' Panel C shows example questions that can be refreshed using Ctrl+I, including 'What's the population of Florida?', 'Is there a pattern in this map?', and 'Which state has the highest population density?' Panel D shows the system's response to 'What else can you do?' listing seven capabilities: compare and sort data, filter data, find similar values and outliers, describe patterns on the map, describe the legend, describe state shapes, and identify neighboring states.
    }
\end{figure*}

\subsubsection{Interactive Map}
To start map interactions, users can press \simplekey{Ctrl + M} and then \simplekey{Tab} to focus on a state, and use arrow keys to move between states.
Since the four cardinal directions are insufficient for all spatial relationships (\textit{e.g.,} both Ohio and Indiana are north of Kentucky), we developed an algorithm that identifies adjacent states, determines cardinal directions by comparing centroids, and selects the closest neighbor in each direction.
For states with complex geometries and atypical centroid positions, we manually adjusted them to ensure natural navigation (\textit{e.g.,} New York, DC, Rhode Island). 
When users attempt to navigate in a direction where no neighboring state exists, the system indicates the boundary condition, \textit{e.g.,} \sayit{There is no state south of Texas}. 
When focused on a state, users can press \simplekey{\texttt{+}} to zoom to county-level. 
The system automatically focuses on the county closest to the state's centroid, with similar arrow key navigation at the county level. 
Users can return to the state-level view using the \simplekey{\texttt{-}} key (\autoref{fig:map-interactions}).

\subsubsection{AI Chat}
The chat interface follows conversational UI design standards~\cite{langevin_heuristic_2021} while accommodating screen-reader interaction.
Users can toggle focus between map and chat using \simplekey{Ctrl + M}. 
Upon focusing on the chat, \sysname announces, \sayit{Type your question here, press enter to submit}. 
After submission, the system repeats the user query followed by a status indicator (\sayit{Looking for answers...}). 
Users can navigate to previous conversation history using \simplekey{Tab}. 
\simplekey{Ctrl + L} repeats the most recent system response while maintaining the input field focus.

On load, \sysname provides an introduction and three selectable example geovisualization questions. 
Below these examples, a \textit{More Suggestions} button (accessible via \simplekey{Ctrl + I}) refreshes the question set, cycling through a list of 12 predefined example questions. 
We also encourage users to ask contextual questions that extend beyond the immediate dataset, \textit{e.g.,} \sayit{What is a choropleth map?}. 
For additional assistance, users can ask \sayit{What else can you do?} to retrieve supported query types. 
\simplekey{Ctrl + H} displays a comprehensive list of shortcuts (\autoref{fig:chat-interactions}).

\subsubsection{Map and Chat Synchronization}
\begin{figure*}[t]
    \centering
    \includegraphics[width=\linewidth]{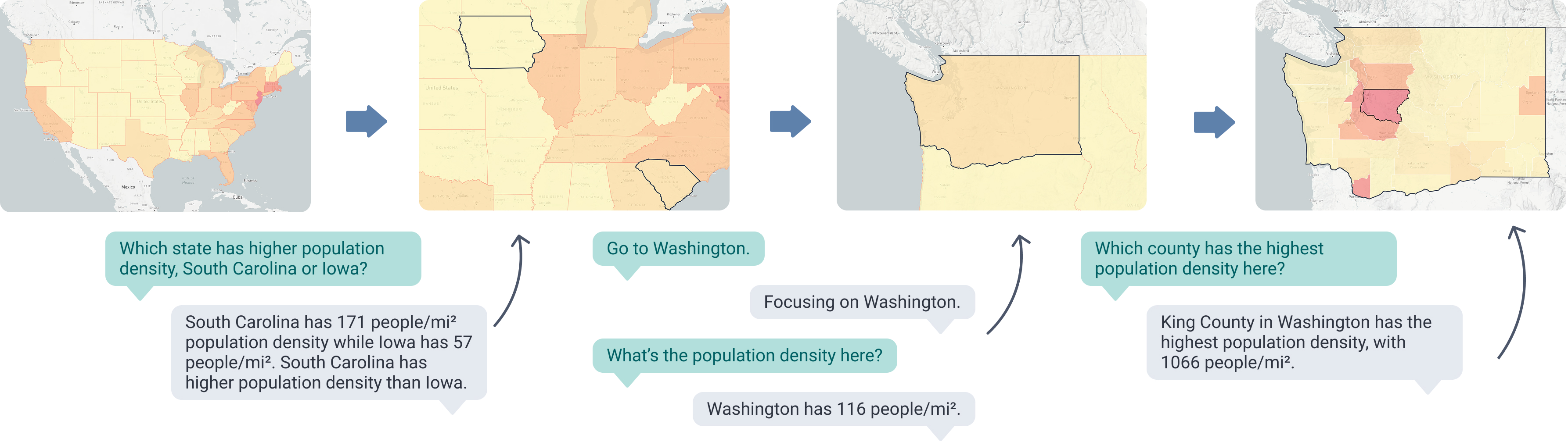}
    \caption{
    Bidirectional synchronization between map and chat interface.
    (A) Comparative queries automatically adjust map zoom to display all referenced states, with textual responses providing enumerated values and answers.
    (B) Users navigate via natural language commands and ask location-specific questions using deictic references like "here."
    (C) Users can zoom to county level and query detailed local information.
    }
    \Description{
    Three-panel sequence demonstrating bidirectional synchronization between map and chat interfaces. Panel 1: User asks 'Which state has higher population density, South Carolina or Iowa?' The system automatically zooms from full US map to regional view showing both states, responding that South Carolina has 171 people/mi² while Iowa has 57 people/mi². Panel 2: User commands 'Go to Washington' and asks 'What's the population density here?' The map focuses on Washington state with status 'Focusing on Washington,' and system responds 'Washington has 116 people/mi².' Panel 3: User asks 'Which county has the highest population density here?' The system zooms to county-level view of Washington with King County highlighted in red, responding 'King County in Washington has the highest population density, with 1066 people/mi².
    }
    \label{fig:sync}
\end{figure*}

To support a tightly integrated, holistic interactive experience, the map and chat components are bidirectionally synchronized. When users query specific geographic entities through AI chat, the system automatically updates the map to provide relevant spatial context. 
For example, when a user asks about a single state's value (\textit{e.g., }\sayit{What is the population density of Illinois?}), the map centers on Illinois and highlights the boundary (\autoref{fig:sync}). 
Beyond  queries, \sysname also responds to explicit navigation commands in natural language, such as \sayit{Take me to Wyoming}, \sayit{Focus on Cook County, Illinois}, or \sayit{Go to Sacramento}, by immediately updating the map focus accordingly. 
Moreover, \sysname maintains the context of the user's current map focus during free exploration, enabling implicit geographic referencing in AI Chat. 
For example, when focused on Colorado, users can ask location-specific questions such as \sayit{What's the population density here?}, \sayit{What are the neighboring states?}, or \sayit{What's the shape of this state?} 
(\autoref{fig:sync}).

\subsection{Implementation}
\textbf{Data Preparation.} 
We obtained geographic boundaries for 48 contiguous U.S. states and 3,222 counties from TIGER shapefiles~\cite{bureau_tigerline_2024}.
Our demographic and socioeconomic datasets are drawn from state- and county-level data from American Community Survey five-year estimates~\cite{bureau_2018-2022_2023}.
These datasets require only minimal processing---merging the shapefiles with the demographic datasets---before being stored in DuckDB to enable efficient queries and analysis.
\newline
\textbf{System Architecture.} \sysname's frontend is implemented in MapboxJS and React.js, with a backend built using Python's Flask framework and DuckDB for geospatial data management. The system uses Google Cloud Platform for hosting and MongoDB for user log data storage.
\newline
\textbf{Speech Integration.} Speech input is implemented by integrating the browser's WebAudio API with a server-side pipeline, using OpenAI's Whisper model for accurate speech-to-text conversion.

%% file: tables/table-query-types.tex
\begin{table*}[]
\centering
\small
\resizebox{\linewidth}{!}{%
\begin{tabular}{@{}p{0.2\linewidth}p{0.2\linewidth}p{0.6\linewidth}@{}}
\textbf{Action/Query}                                & \textbf{Detailed Type} & \textbf{Example}                                                        \\ 
\midrule
\rowcolor[HTML]{EFF2F6} 
Map Action & & Go to Boston\\

\\[-0.8em]

\rowcolor[HTML]{F9FBE7} 
\cellcolor[HTML]{F9FBE7}                       & Retrieve               & What's the population density of Vermont?                             \\
\rowcolor[HTML]{F9FBE7} 
\cellcolor[HTML]{F9FBE7}                                   & Compare                & Which state has higher population density, Louisiana or South Dakota? \\
\rowcolor[HTML]{F9FBE7} 
\cellcolor[HTML]{F9FBE7}                                   & Find Extremum          & Which state has the highest population density?                       \\
\rowcolor[HTML]{F9FBE7} 
\cellcolor[HTML]{F9FBE7}                                   & Aggregate              & What's the average population density?                                \\
\rowcolor[HTML]{F9FBE7} 
\cellcolor[HTML]{F9FBE7}                                   & Filter                 & Which states have density over 300 people/sqm?            \\
\rowcolor[HTML]{F9FBE7} 
\cellcolor[HTML]{F9FBE7}                                   & Sort                   & Top 5 states with the highest population density?          \\
\rowcolor[HTML]{F9FBE7} 
\multirow{-7}{*}[.5em]{\cellcolor[HTML]{F9FBE7} Analytical Query} & Cluster                & Which state has a similar population density to Oregon?               \\

\\[-0.8em]

\rowcolor[HTML]{E8F5E9} 
\cellcolor[HTML]{E8F5E9}                                   & Pattern                & Is there a pattern on the map? Can you describe it?                   \\
\rowcolor[HTML]{E8F5E9} 
\multirow{-2}{*}[.1em]{\cellcolor[HTML]{E8F5E9}Geospatial Query} & Outlier                & What are the outliers?                                   \\

\\[-0.8em]

\rowcolor[HTML]{E0F2F1} 
\cellcolor[HTML]{E0F2F1}                                   & Legend                 & Can you tell me more about the legend?                                \\
\rowcolor[HTML]{E0F2F1} 
\cellcolor[HTML]{E0F2F1}                                   & Shape                  & What is the shape of Wisconsin?                                       \\
\rowcolor[HTML]{E0F2F1} 
\multirow{-3}{*}[-.2em]{\cellcolor[HTML]{E0F2F1}Visual Query}     & \mbox{Spatial Relationships}  & What are the neighboring states of Illinois?                          \\

\\[-0.8em]

\rowcolor[HTML]{E8EAF6} 
\cellcolor[HTML]{E8EAF6}                                   & Visualization Knowledge & What is a choropleth map?                                             \\
\rowcolor[HTML]{E8EAF6} 
\multirow{-2}{*}{\cellcolor[HTML]{E8EAF6} Contextual Query} & \mbox{General Knowledge} & Is there a relationship between income and population density? \\ 
\end{tabular}%
}
\caption{Supported query types handled by the Query Processor with example questions.}
\label{tab:query-types}
\end{table*}

%% file: tables/table-shortcut.tex
\begin{table}[b]
\centering
\small
\renewcommand{\arraystretch}{1.2}
\begin{tabular}{@{}p{0.4\linewidth}p{0.5\linewidth}@{}}
\multicolumn{1}{l}{\textbf{Shortcut / Key}} & \multicolumn{1}{l}{\textbf{Function}} \\
\midrule
\rowcolor[HTML]{FFFFFF} 
\simplekey{Ctrl} + \simplekey{M} & Toggle between map and chat interaction \\
\rowcolor[HTML]{EFF2F6} 
\simplekey{{$\leftarrow$}}~~\simplekey{{$\uparrow$}}~~\simplekey{{$\rightarrow$}}~~\simplekey{{$\downarrow$}} & Navigate between states/counties \\
\rowcolor[HTML]{FFFFFF} 
\simplekey{\texttt{+}} & Zoom in to county level within a state \\
\rowcolor[HTML]{EFF2F6} 
\simplekey{\texttt{-}} & Zoom out to state level \\
\rowcolor[HTML]{FFFFFF} 
\simplekey{Ctrl} + \simplekey{L} & Hear the last response again \\
\rowcolor[HTML]{EFF2F6} 
\simplekey{Ctrl} + \simplekey{H} & Show/hide the help window \\
\rowcolor[HTML]{FFFFFF} 
\simplekey{Ctrl} + \simplekey{I} & Refresh question suggestions \\
\end{tabular}
\caption{List of all shortcut keys and associated behaviors.}
\label{tab:shortcut}
\end{table}

%% file: sections/5-user-study.tex
\section{User Study}

To examine user engagement with an accessible geovisualization QA system and compare interaction patterns and query strategies between blind and sighted users (RQ1--4), we conducted a user study with both groups.
As the first AI-based accessible geovisualization QA system, our study goals were twofold: first, to explore what \textit{types of queries} do screen-reader users make when interacting with an accessible geovisualization; and second, to examine \textit{how well} the current \sysname prototype supports these queries, builds appropriate mental models of the underlying data, and leads to accurate insights and takeaways. 
All study sessions were audio and video recorded. 
We used a combination of qualitative and quantitative measures to address our research goals.

\subsection{Participants}

We recruited six screen-reader users (referred to as B1-B6) through mailing lists and six sighted researchers and practitioners in human-computer interaction, computer science, and urban planning, with varying levels of U.S. geography knowledge (S1-S6).
Before the study, participants filled out a questionnaire to record their demographic information, education level, daily computer usage, frequency of interaction with map visualizations, AI chatbots, and voice assistants. 
Screen-reader users answered additional questions about their screen-reader software and vision-loss level (\autoref{tab:participant}).

\begin{figure*}[b]
    \centering
    \includegraphics[width=\linewidth]{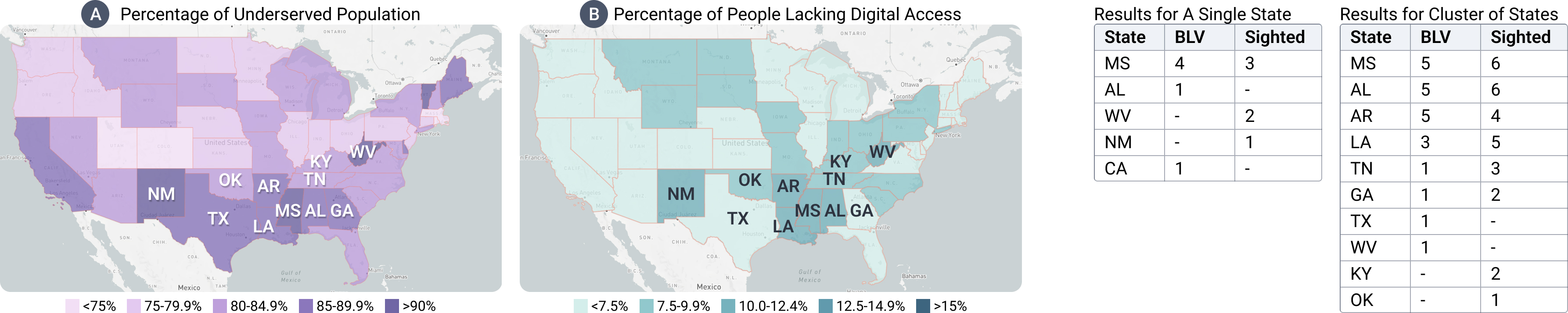}
    \caption{Task 1 visualizations and results. 
    (A) Choropleth map showing the percentage of underserved population across states. 
    (B) Choropleth map showing the percentage of people lacking digital access.  
    Tables present participant funding choices for both a single state and a cluster of states, revealing convergence in pattern identification between BLV and sighted participants.}
    \Description{
    Task 1 visualization results showing two choropleth maps and participant data tables. Panel A displays a choropleth map of percentage of underserved population across US states, with darker purple indicating higher percentages (legend: <75\%, 75-79.9\%, 80-84.9\%, 85-89.9\%, >90\%). States like Texas, New Mexico, and several southeastern states show darker shading. Panel B shows a choropleth map of percentage of people lacking digital access, with darker blue-green indicating higher percentages (legend: <7.5\%, 7.5-9.9\%, 10.0-12.4\%, 12.5-14.9\%, >15\%). States like New Mexico, Arkansas, and West Virginia show darker shading. Two data tables show participant funding choices: 'Results for A Single State' lists individual states (MS, AL, WV, NM, CA) with BLV and Sighted participant counts, while 'Results for Cluster of States' shows the same states with different participant distributions. The data reveals convergence in pattern identification between blind/low vision (BLV) and sighted participants.
    }
    \label{fig:task1}
\end{figure*}

\subsection{Study Design}
The core of our study comprised two data analysis tasks.
We intentionally designed tasks requiring complex, multi-step analytical processes rather than single-question interactions.
To create a natural exploratory experience and ensure participant engagement, we based our tasks on existing data stories and real-life scenarios.
Prior to the study sessions, we conducted pilot studies with co-design participants and one sighted member of the research team.

In Task 1, participants were asked to imagine themselves as decision-makers responsible for distributing \textit{State Digital Equity Planning Grant} funding across states---a task adapted from the U.S. Census on mapping digital equity, which originally contained inaccessible geovisualizations\footnote{https://www.census.gov/library/stories/2022/05/mapping-digital-equity-in-every-state.html}.
Participants were asked to select both a single state and a cluster of four to six geographically contiguous states most deserving of funding allocation.

\begin{figure*}[t]
    \centering
    \includegraphics[width=\linewidth]{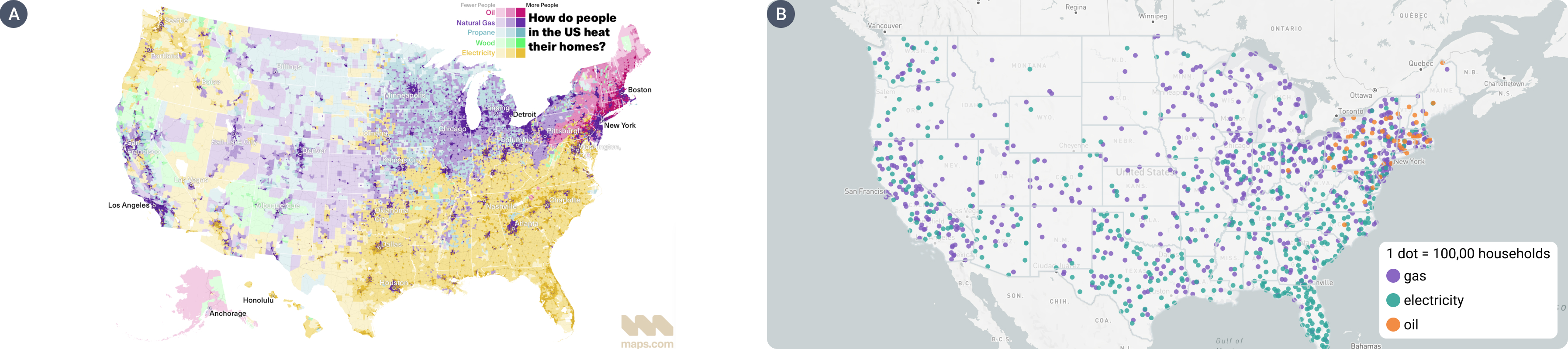}
    \caption{
    Task 2 comparison of (A) original Washington Post choropleth map showing predominant home heating fuel sources and (B) our adapted dot density version.
    The dot density map removes the immediate visual patterns available in the choropleth map, allowing us to explore how well our system could support sighted users when immediate visual patterns are unavailable.
    }
    \Description{
    Task 2 comparison of two visualization approaches for US home heating fuel data. Panel A shows the original Washington Post choropleth map with distinct color regions indicating predominant heating fuel sources across states, displaying clear visual patterns of regional clustering. Panel B shows the same data converted to a dot density map where each dot represents 100,000 households, with blue dots for gas, purple dots for electricity, and orange dots for oil heating. The dots are scattered across the US without obvious regional color clustering, removing the immediate visual patterns visible in the choropleth version. This design tests how well the accessible system supports sighted users when geographic patterns require deeper analysis rather than immediate visual recognition.
    }
    \label{fig:task2}
\end{figure*}

For Task 2, we adapted a \textit{Washington Post} article on U.S. home heating\footnote{https://www.washingtonpost.com/climate-environment/interactive/2023/home-electrification-heat-pumps-gas-furnace/}. 
To explore how well our system supports sighted users when immediate visual patterns were unavailable, we transformed the original choropleth maps into dot density maps.
Each dot represented 100K households and was randomly distributed across its corresponding geographic region (\autoref{fig:task2}).
Participants were asked to identify  predominant heating fuel sources in different U.S. regions and to explain these patterns.

\input{tables/table-participant}

\subsection{Procedure}
Study sessions were conducted individually via Zoom, with each session lasting between 90 and 120 minutes. Every session followed a four-part protocol.
\newline
\textbf{Introduction.}
Sessions began with an introduction to the project and an overview of the study agenda. 
After obtaining informed consent, we asked participants about their experience with AI chatbots and geovisualizations.
\newline
\textbf{Tutorial \& Free Exploration.}
The researcher then guided participants through a tutorial using the U.S. population density dataset, covering \sysname's capabilities: answerable question types, keyboard navigation, zoom controls, shortcut keys, and voice input.
Participants were then encouraged to freely explore the system.
\newline
\textbf{Study Tasks.}
Participants were given up to 30 minutes to complete the two analytical tasks.
As an exploratory study focused on analytical approaches rather than completion rates, we allowed participants sufficient time to explore and demonstrate reasoning.
If the combined tutorial, exploration, and Task 1 phases exceeded one hour, participants were not asked to proceed to Task 2, ensuring adequate time remained for the interview and debrief.
For both tasks, we emphasized our interest in studying participants' analytical processes rather than their derived answers.
When researchers observed immediate inaccuracies in system responses during the session, they notified participants to ensure the study could continue effectively and to prevent participants from building an analysis on incorrect information.
\newline
\textbf{Semi-structured Interview \& Debrief.}
Following task completion, the researcher conducted semi-structured interviews to gather qualitative insights about participants' experiences, challenges, and suggestions for improvement.
At the conclusion of each session, participants received a comprehensive debrief explaining the technical architecture of \sysname and were shared the original sources of the data stories upon request.

\subsection{Evaluation Measures}
We used both objective and subjective evaluation measures to assess system performance. 
The objective measure consisted of the percentage of questions that the system answered correctly during user evaluation.
We determined correctness by comparing system responses against ground truth data from the original datasets (American Community Survey) and verified geographic information.
For interpretive questions, two researchers independently assessed answer accuracy based on whether responses were factually supported by the underlying data.
In cases of disagreement, the two researchers discussed their assessments until reaching agreement.

For subjective evaluation, we adapted measures from prior work on map visualization usage: \textit{reading, analysis}, and \textit{interpretation}~\cite{muehrcke_functional_1978}. 
Given our focus on interactive maps, we also included \textit{navigation} as a fourth dimension. Additionally, we evaluated chat functionality and the relative importance of both map and chat components. All subjective measures were captured using 7-point Likert scale questions followed by open-ended questions (\autoref{fig:likert-scale}). 

\subsection{Data and Analysis}
All study sessions were audio and video recorded. 
Our analysis focused on summarizing high-level themes through qualitative open coding~\cite{charmaz_constructing_2006}. 
One researcher developed a set of themes based on video transcripts and observational notes, then thematically coded the responses.
A second researcher verified these themes~\cite{spall_peer_1998}.
For system performance, two researchers independently validated all answers; disagreements were discussed until consensus.

%% file: tables/table-participant.tex
\begin{table*}[t]
\centering
\small 
\renewcommand{\arraystretch}{1.2}
\begin{tabular}
{@{}>{\raggedright\arraybackslash}p{0.5cm}>{\raggedright\arraybackslash}p{0.8cm}>
{\raggedright\arraybackslash}p{0.8cm}>
{\raggedright\arraybackslash}p{1.2cm}>
{\raggedright\arraybackslash}p{1.4cm}>{\raggedright\arraybackslash}p{1.4cm}>{\raggedright\arraybackslash}p{1.4cm}>{\raggedright\arraybackslash}p{1.9cm}>{\raggedright\arraybackslash}p{1.8cm}>{\raggedright\arraybackslash}p{2.4cm}@{}}
\rowcolor[HTML]{FFFFFF} 
\textbf{PID} &
  \textbf{Age} &
  \textbf{Gender} &
  \textbf{Education Level} &
  \textbf{Computer Usage} &
  \textbf{Online Map Usage} &
  \textbf{AI Chatbot Usage} &
  \textbf{Voice Assistant Usage} &
  \textbf{Screen-reader} &
  \textbf{Vision-loss Level} \\ \midrule
\rowcolor[HTML]{EFF2F6} 
C1 & 25-34 & Woman   & Masters   & 5 hours+  & Occasionally       & Occasionally       & Daily     & NVDA                 & Blind since birth                         \\
\rowcolor[HTML]{FFFFFF} 
C2 & 45-54 & Man & Masters   & 3-5 hours  & Frequently       & Occasionally      & Daily           & JAWS                 & Lost vision gradually, Complete blindness \\ 
\rowcolor[HTML]{EFF2F6} 
B1 & 25-34 & Man   & Bachelor   & 5 hours+  & Rarely       & Regularly       & Occasionally    & NVDA                 & Blind since birth                         \\
\rowcolor[HTML]{FFFFFF} 
B2 & 55-64 & Woman & Bachelor   & 5 hours+  & Rarely       & Regularly       & Daily           & JAWS                 & Lost vision gradually, Complete blindness \\
\rowcolor[HTML]{EFF2F6} 
B3 & 55-64 & Man   & Masters     & 5 hours+  & Regularly    & Daily           & Daily           & JAWS                 & Lost vision gradually                     \\
\rowcolor[HTML]{FFFFFF} 
B4 & 35-44 & Woman & High school & 0-1 hours & Occasionally & Never used & Daily           & JAWS/Voice Over      & Complete blindness                        \\
\rowcolor[HTML]{EFF2F6} 
B5 & 55-64 & Man   & Bachelor   & 5 hours+  & Frequently   & Frequently      & Frequently      & JAWS/Voice Over/NVDA & Lost vision gradually, Complete blindness \\
\rowcolor[HTML]{FFFFFF} 
B6 & 25-34 & Man   & High school & 5 hours+  & Occasionally & Daily           & Daily           & NVDA                 & Blind since birth                         \\
\rowcolor[HTML]{EFF2F6} 
S1 & 25-34 & Woman & Masters     & 5 hours+  & Occasionally & Daily           & Rarely          &                      &                                           \\
\rowcolor[HTML]{FFFFFF} 
S2 & 25-34 & Woman & Masters     & 5 hours+  & Occasionally & Regularly       & Frequently      &                      &                                           \\
\rowcolor[HTML]{EFF2F6} 
S3 & 25-34 & Man   & Doctoral & 5 hours+  & Daily        & Daily           & Never used &                      &                                           \\
\rowcolor[HTML]{FFFFFF} 
S4 & 25-34 & Man   & Doctoral & 5 hours+  & Frequently   & Frequently      & Occasionally    &                      &                                           \\
\rowcolor[HTML]{EFF2F6} 
S5 & 18-24 & Woman & Bachelor & 5 hours+  & Occasionally & Daily           & Never used &                      &                                           \\
\rowcolor[HTML]{FFFFFF} 
S6 & 25-34 & Man   & Masters     & 5 hours+  & Occasionally & Frequently      & Frequently      &                      &                                           \\ 
\end{tabular}%
\vspace{1em}
\caption{Participant demographics and technology experience.  
C1-C2 are co-design participants, B1-B6 are screen-reader users, S1-S6 are sighted participants. 
Computer usage is based on daily usage hours.
For all other technology usage, frequently = multiple times per week, regularly = weekly, occasionally = few times a month, rarely = less than once a month.}
\label{tab:participant}
\vspace{-0.5cm}
\end{table*}

%% file: sections/6-findings.tex
\section{Findings}

Overall, participants reacted positively to the \sysname prototype.
BLV participants particularly valued the system's clear and detailed spatial descriptions, the ability to navigate maps independently using natural language commands, and access to analytical capabilities previously unavailable in accessible geovisualization tools.
Half of the BLV participants requested access to \sysname for continued use after the study session.

All participants completed at least one study task.
On average BLV participants spent 30.9 minutes ($SD=12.2$) interacting with the tutorial site,  $18.2$ minutes ($SD=5.8$) on Task 1, and 17.3 minutes ($SD=0.75$) on Task 2.
Sighted participants spent $18.6$ minutes ($SD=5.4$) interacting with the tutorial site, then 14.2 minutes ($SD=4.5$) on Task 1 and 19.0 minutes ($SD=5.3$) on Task 2.
Below we organize our findings around our research questions.

\subsection{Types of Query Asked \& System Performance (RQ1)}
A key function of \sysname is to pose natural language geoanalytic questions or enact UI commands via the chat interface. 
We begin by analyzing the types of queries posed and how well \sysname responded.

\textbf{Types of Query.}
Our participants ($N=12$) asked \sysname a total of 346 questions.
~\autoref{tab:study-question-types} details the query type distribution by user group.
Both BLV and sighted users shared the same top three most frequent query categories.
These were \textit{general knowledge} queries (\textit{e.g.,} 
\sayhi{Why do Midwestern states use predominantly gas?}), followed by \textit{pattern} queries (\textit{e.g.,} 
\sayhi{Is there a pattern in this region regarding broadband access?}), and \textit{action} queries (\textit{e.g.,} \sayhi{Take me back to Arizona}).
Usage patterns then diverged.  
BLV users most frequently asked \textit{retrieval} (\textit{e.g.,} \sayhi{How many households use gas heating in Texas? }) and \textit{find extremum} (\textit{e.g.,} \sayhi{Which state in the United States has the highest percentage of people lacking broadband access? }) queries, while sighted users favored \textit{aggregate} (\textit{e.g.,} \sayhi{Give me the population density for each of these and the standard deviation compared to the mean.}) and \textit{sort} (\textit{e.g.,} \sayhi{Tell me the top 5 states lacking broadband access}) queries.
As expected, \textit{visual queries} (shape and spatial relationship) were asked almost exclusively by BLV users (\textit{e.g.,} \sayhi{What's the shape of Alabama?} and \sayhi{What are Florida's neighboring states?}).

Impressively, we found that \sysname supported 92\% of the total questions asked.
The remaining 8\% fell into two categories. 
2\% of the questions were outside of our defined query types (\autoref{tab:study-question-types}); these included requests to create additional visualizations (\textit{e.g.,} \sayhi{Can you highlight the six states on the map?}) or to recolor visualizations based on different data (\textit{e.g.,} \sayhi{Can you color the states based on the most common fuel source?}).
The other 6\% were correctly classified into supported query types (\textit{e.g.,} \textit{Aggregate}), but \sysname could not provide an answer because the required calculation was not yet implemented in our backend (\textit{e.g.,} \sayit{Give me the population density for each of these and the standard deviation compared to the mean}).

\input{tables/table-study-question-types}

\textbf{Accuracy.}
Because \sysname incorporates statistical analyses, database queries, and LLM-based analytics, evaluating response accuracy is critical~\cite{pang_understanding_2025}. 
In total, 289 (83.8\%) questions were answered correctly.
Of the 56 incorrect responses, we identified specific failure points within our pipeline (\autoref{tab:error-types}). Query Refiner made 15 (26.8\%) mistakes, primarily failing to retrieve the current user focus or relevant chat history for disambiguation. For example, when a user focused on Washington state and asked \sayhi{What's the population density here?}, the system returned a generic definition of population density: \sayhi{To determine population density, divide the total population of an area by its land area (in square miles or kilometers)...} 
Scope Assessor made 12 (21.4\%) mistakes, incorrectly routing questions either to GPT when they should have been processed locally or to local processing when they required external knowledge. 
For instance, when asked \sayhi{What is the percentage of underserved populations here?}, the system incorrectly determined this to be out of scope and provided a generic definition rather than analyzing the available demographic data.
Query Processor made 3 (5.4\%) classification errors, such as misclassifying \sayhi{Why do rural counties use more electricity than gas?} as a comparison query rather than general knowledge, resulting in data comparisons instead of explanatory content. The remaining 26 (46.4\%) incorrect responses stemmed from limitations beyond our current system capabilities, including requests for visualization creation (e.g., \sayhi{Can you create a heat map for the gas heating fuel type layer?}), GPT hallucinations (such as incorrectly claiming geographic connections), ambiguously phrased user questions (e.g., incomplete queries like \sayit{states with lower population density}), and server errors.

\subsection{Geovisualization Engagement (RQ2 \& RQ3)}
Here we analyze how both BLV and sighted participants engaged with \sysname across four categories of geovisualization usage: reading, analysis, interpretation, and navigation. 

\begin{figure}[b]
    \centering
    \includegraphics[width=\linewidth]{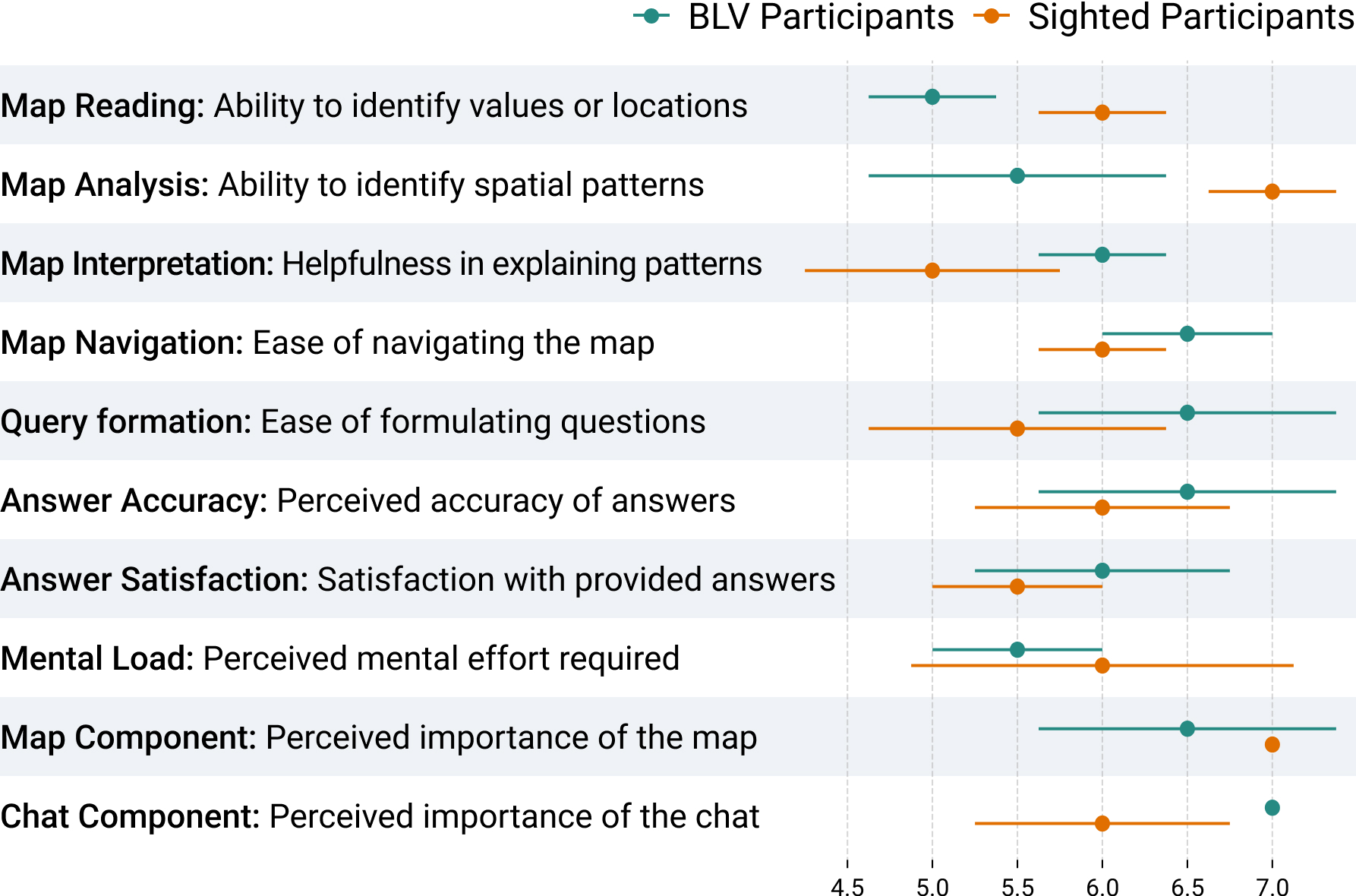}
    \caption{Likert scale ratings (1-7) across multiple dimensions, where 7 represents better outcomes (\textit{i.e.,} easier, more accurate, more important). Plot shows median values with IQR for each user group. BLV participants unanimously rated the chat component as 7/7, while sighted participants unanimously gave the map component 7/7.}
    \Description{
    Likert scale rating comparison (1-7 scale, where 7 represents better outcomes such as easier, more accurate, more important) between BLV and sighted participants across multiple evaluation dimensions. Plot displays median values with interquartile range (IQR) bars for each user group. Key finding: BLV participants unanimously rated the chat component as 7/7, while sighted participants unanimously gave the map component 7/7, highlighting different interface preferences between the two user groups.
    }
    \label{fig:likert-scale}
\end{figure}

\textbf{Map Reading.}
When asked to rate the ease of identifying specific values or regions on the map (on a scale of 1-7), BLV participants reported a median of 5 (\textit{IQR} = 0.75), while sighted participants reported a median of 6 (\textit{IQR} = 0.75) (\autoref{fig:likert-scale}). For BLV participants, \sysname effectively supported basic map reading tasks while helping recall their geographic knowledge. 
As B3 noted: \sayhi{This is really neat ... because sometimes, I'll go through all 50 states in my head, and I can't remember things like, `Oh, Rhode Island is south of Massachusetts.'}
Regarding the map-reading answer quality, BLV participants described the system's responses as \sayhi{very informative} (B5), \sayhi{clear, and concise} (B5).
However, some (B3, B4) raised concerns about verifying the provided information.
As B4 noted: \sayhi{I'd probably ask Alexa and see if their answers match; if they do for a couple of questions, I would trust [GeoVisA11y] more.}
Beyond data-related queries, participants appreciated \sysname's ability to describe state shapes and geographic relationships, with B6 commenting \sayhi{It really describes things well, like the shape of Alabama.}

\input{tables/table-error-types}

Sighted participants approached map reading differently, valuing how \sysname prevented misinterpretation through natural language queries. 
S1 observed: \sayhi{I think being able to ask natural questions really helps because otherwise I would have thought that 5\% represents the population that has broadband access, which is incorrect.} 
Having both the map and the chat let participants verify information across channels; as S6 mentioned, \sayhi{I was able to corroborate information given by the chat with the map.}

Both user groups benefited from \sysname's highlighting function, though for different reasons. 
For sighted users with limited U.S. geographic knowledge, the feature provided immediate orientation: \sayhi{I like that it highlighted the different states related to the question, because I don't really know where New Jersey is} (S2). 
For BLV users, highlighting served as a trigger for screen-reader focus on the map, creating a multimodal signal of spatial information.

\textbf{Map Analysis.}
When asked to rate the ease of identifying spatial trends and patterns on the map (on a scale of 1-7), BLV participants reported a median of 5.5 (\textit{IQR} = 1.75), while sighted participants reported a median of 7 (\textit{IQR} = 0.75) (\autoref{fig:likert-scale}).
BLV participants appreciated the system's ability to translate complex spatial data into accessible language. 
B2 noted \sayhi{It uses language that's not complicated. It's very general, easy to understand.} 
B5 highlighted \sysname's analytical advantage over other tools: \sayhi{This is much better than most maps that I've seen, because you can't ask those kind of questions and get that feedback. Mostly, ... [other maps are] just a superficial layer of the map.} 

Sighted participants similarly valued the system's analytical capabilities, but with greater emphasis on its dynamic visualization features. 
S3 appreciated that \sayhi{it automatically prepares these additional visualization layers to help me [identify patterns],} while S5 noted that \sayhi{the coolest thing is that you can add layers to help you visualize certain patterns.} 
S1 noted \sysname's advantage compared to accessing only the visualization: \sayhi{If I only had the map, the information will all be scattered, and I'll need to manually collect them, note them down, then compare the numbers since the color doesn't really show much when the numbers are very close.} 

\textbf{Map Interpretation.}
When asked to rate how well the system helped them understand why certain patterns or relationships exist (on a scale of 1-7), BLV participants reported a median of 6 (\textit{IQR} = 0.75), while sighted participants reported a median of 5 (\textit{IQR} = 1.5) (\autoref{fig:likert-scale}).
B5 expressed how \sysname facilitated engaging pattern-finding: \sayhi{What I enjoyed the most was the ability to query across datasets, to find relevancy from the information within each and come up with an answer that gave me more specific targets.}
However, they sometimes requested more specific explanations: \sayhi{I remember asking the system to explain the [heating fuel] pattern, and it mentioned climate factors. If we're able to get specific data like average temperatures for different regions, it would provide me an immediate snapshot of why those patterns exist} (C2).

Sighted participants demonstrated diverse approaches to map interpretation based on their background knowledge.
S4 relied on the system for explanations after identifying the patterns independently, asking questions like, \sayhi{Why does Washington use more electricity than gas for heating homes?}
While S5 used the system for pattern identification, then drew on personal expertise for interpretation, stating, \sayhi{Ah, I can definitely think of the potential reasons for this. In Washington, electricity is generated from various renewable sources, like hydropower and wind.}
The differences in interpretation strategies highlight how \sysname accommodated diverse knowledge levels, providing necessary context for users with limited domain expertise (both BLV and sighted) while facilitating more efficient pattern identification for users who could then interpret findings using their specialized knowledge.

\textbf{Map Navigation.}
When asked to rate their navigation experience, BLV participants reported a median of 6.5 (\textit{IQR} = 1), and sighted participants reported a median of 6 (\textit{IQR} = 0.75).
BLV participants highlighted the unprecedented autonomy offered by \sysname's navigation: \sayhi{This is pretty cool because I have never understood mapping or tactile graphs this way before. When you're studying history, you're always feeling tactile graphs just to fill in the blanks or feel what the teachers ask you to. But now I can explore with free will, using the arrow keys to see for myself} (B1). 
Some participants reported that navigation triggered their latent spatial knowledge: \sayhi{Oh, wow. I just realized what I did because of my knowledge of the Northeast. I arrowed around because I know Vermont is pretty close to New York} (C2).

Navigation preferences differed between groups, with BLV participants primarily using keyboard navigation and sighted participants preferring direct mouse interaction.
As S6 explained: \sayit{I was expecting that I would be able to click on the states as well because I'm more of a mouse user than a keyboard user.}
Despite these different interaction preferences, both groups particularly valued the semantic navigation commands.
B6 mentioned: \sayhi{Being able to tell it where I wanted to go, like 'go to Georgia,' that's not something most programs use. It's really nice.} (B6), while sighted users found it helpful when dealing with unfamiliar geography: \sayhi{I think this go-to function is very helpful because I don't actually know where the states are given my limited knowledge of U.S. geography} (S2). 

\subsection{Task Completion \& Query Strategies (RQ4)}
Here we analyze how BLV and sighted participants completed Task~1 (distributing federal broadband funding) and Task~2 (identifying regional heating fuel patterns), focusing on their querying approaches.
Due to time constraints, only two BLV participants completed Task~2, so our quantitative analysis focuses on Task~1, with qualitative insights drawn from Task~2 where relevant.

\textbf{BLV Participants.} Overall, five of the six BLV participants came up with reasonable answers for Task 1, completing the task in an average of 18.2 minutes ($SD = 5.8$ minutes).
For individual state funding recommendations, participants selected Mississippi and Alabama.
For the cluster funding recommendation, participants identified Mississippi, Arkansas, Alabama, Louisiana, Texas, Georgia, West Virginia, and Tennessee (\autoref{fig:task1}).
One participant insisted on funding California despite the data, explaining it was their home state.
BLV participants employed diverse query strategies, including finding extremum, sorting, comparing, and inquiring about geographic patterns and spatial relationships.
They used a combination of verbal queries and key navigation to determine spatial relationships between target states. 

Participants demonstrated adaptability when encountering system misinterpretations. 
B5 initially used ambiguous pronouns beyond the system's capabilities by asking \sayhi{What are \textbf{my} neighboring states?}. 
When they recognized the system struggled to answer such a question, they self-corrected by specifying state name.
When B2 asked about \sayhi{fuel patterns} and received information about transportation fuels, they refined their query to specify \sayhi{heating fuels for homes}. 
In another instance, B3 requested \sayhi{Select a group of six contiguous states that have low access to broadband,} which the system incorrectly answered \sayhi{Mississippi, Arkansas, Louisiana, West Virginia, Kentucky, Alabama}. 
The participant recognized the geographical error, noting \sayhi{West Virginia isn't connected right?} and then verified using keyboard navigation to confirm the mistake.

Notably, BLV participants demonstrated greater reliance on the researcher's task descriptions when formulating queries and higher expectations of the system's analytical capabilities. 
For example, B3 directly asked: \sayhi{Which state do you think funding should be appropriated to based on the percentage of lack of broadband access?}

\textbf{Sighted Participants.}
All six sighted participants developed reasonable answers for Task 1, recommending Mississippi, West Virginia, and New Mexico as individual states, and identifying a southeastern cluster including Mississippi, Alabama, Louisiana, and Arkansas as priority regions (\autoref{fig:task1}).
Comparing results from BLV participants with sighted participants reveals a general convergence in pattern identification between the two groups. 
For single state selection, both BLV and sighted participants identified Mississippi most frequently. 
When analyzing clusters of states, Mississippi, Alabama, Louisiana and Arkansas were chosen most frequently in both groups. The results begin to diverge for secondary states, with BLV participants selecting Tennessee, Georgia, Texas, and West Virginia (each chosen once), while sighted participants identified Tennessee (3), Georgia (2), Kentucky (2), and Oklahoma (1).

Although sighted participants used similar query types (finding extremum, sorting, comparing, and geographic patterns), they visually narrowed selections and determined geographical connectivity before making specific queries.
Their system usage varied based on domain expertise and map complexity. 
S4 (proficient with gaming visualizations) identified patterns independently but used the system for interpretations, while S5 (urban planning researcher) primarily applied their domain knowledge after using the system to locate patterns. 
Geographic familiarity influenced interaction approaches—S1 (unfamiliar with U.S. geography) relied heavily on text commands like \sayit{Go to Vermont}. 
All sighted participants demonstrated increased chat usage when working with dot density maps compared to choropleth maps.

While the system could not answer certain questions from BLV participants due to ambiguity, most limitations encountered by sighted participants were related to requests for expanded capabilities.
Sighted participants frequently requested enhanced visualization features, such as \sayhi{Can you plot the size of each state?} (S3) and \sayhi{Can you make the dots bigger?} (S5). 
Additionally, sighted users requested more sophisticated statistical analysis, including standard deviation calculations and percentile-based rankings.

\subsection{Suggested Improvements \& Applications.}
We next summarize participant feedback on potential system enhancements for the map component, the chat component, and the synchronization between these two. 

\textbf{Enhanced Navigation.}
BLV participants frequently ($N=4/6$) requested improved navigation capabilities.
B1 and B2 specifically cited navigation as both their most enjoyable and most challenging aspect of the system. 
The current implementation supports only cardinal direction navigation, which falls short for the complex spatial relationships. 
For instance, moving south from Vermont directs users to Massachusetts, but navigating back north also directs them to New Hampshire (since both Vermont and New Hampshire border Massachusetts to the north, with New Hampshire's centroid closer to Massachusetts's centroid). 
This asymmetric directionality created confusion for BLV participants unfamiliar with the intricate spatial relationships between states.

\textbf{More Specific, Structured Answers with Source Attribution.} 
Participants from both groups observed that answers related to map interpretation lacked specificity (S1, S4, B2). 
Unstructured responses or responses structured differently than previous responses made participants question answer accuracy, as S1 noted, \sayhi{I paused a little and wondered is this part of information that's hallucinated?} B3 suggested incorporating hyperlinks along with answers to facilitate verification of information sources.

\textbf{More Transparent Reasoning Processes.} 
When posing higher level synthesis questions requiring integration across multiple datasets, participants wanted more transparency into the analytical processes. 
S3 noted that the system \sayhi{doesn't explicitly say why it made certain decisions}, and S6 added, \sayhi{I was trying to understand why the system would consider certain counties to be outliers.}

\textbf{Improved Map-Chat Synchronization.}
S2 identified a mismatch between the map and the chat: \sayhi{Sometimes, there's a mismatch between what I think the chat is selecting versus what the map is actually focusing on.} 
This mismatch occurred when \sysname couldn't answer using local data, resulting in GPT responses that didn't trigger a geographic unit on the map.
S1 wanted more direct interaction between the visual and conversational interfaces: \sayhi{I wish I can just click on the map to highlight four states and then ask what are the average income of these four.}

\textbf{Potential Applications.}
Although we did not explicitly solicit input on potential applications, BLV participants spontaneously identified various use cases, including interpreting election maps, accessing supplementary information after listening to the news, and understanding resource distribution following natural disasters. 
B5 drew on professional experience to suggest applications: \sayhi{I worked for [a state government department]. There is an audience for this application in public service, especially among those who are in charge of distributing funds or assessing taxes.} 
B6 highlighted the educational value: \sayhi{This could be really good for college students who are going into any kind of social sciences.}

%% file: tables/table-study-question-types.tex
\begin{table*}[t]
\centering
\small 
\renewcommand{\arraystretch}{1.2}
\begin{tabular}
{@{}lrrrrrrp{8cm}@{}}
\textbf{Query Type} &
  \multicolumn{1}{l}{\textbf{BLV}} &
  \multicolumn{1}{l}{\textbf{\%}} &
  \multicolumn{1}{l}{\textbf{Sighted}} &
  \multicolumn{1}{l}{\textbf{\%}} &
  \multicolumn{1}{l}{\textbf{Total}} &
  \multicolumn{1}{l}{\textbf{\%}} &
  \textbf{Example} \\ \midrule
Action &
  27 &
  \cellcolor[HTML]{81CCC5}18.0\% &
  29 &
  \cellcolor[HTML]{9AD6D0}14.9\% &
  56 &
  \cellcolor[HTML]{8FD2CB}16.2\% &
  \sayit{Take me back to Arizona.} (B4) \\
Retrieve &
  22 &
  \cellcolor[HTML]{98D5CF}14.7\% &
  10 &
  \cellcolor[HTML]{DDF1EF}5.1\% &
  32 &
  \cellcolor[HTML]{BFE5E2}9.3\% &
  \sayit{How many households use gas heating in Texas?} (B2) \\
Compare &
  4 &
  \cellcolor[HTML]{EDF8F7}2.7\% &
  4 &
  \cellcolor[HTML]{F2FAF9}2.1\% &
  8 &
  \cellcolor[HTML]{EFF9F8}2.3\% &
  \sayit {Compare OK and NE populations.} (B5) \\
Find Extremum &
  12 &
  \cellcolor[HTML]{C7E8E5}8.0\% &
  9 &
  \cellcolor[HTML]{E0F3F1}4.6\% &
  21 &
  \cellcolor[HTML]{D5EEEC}6.1\% &
  \sayit{Which state in the United States has the highest percentage of people lacking broadband access?} (B3) \\
Aggregate &
  3 &
  \cellcolor[HTML]{F1FAF9}2.0\% &
  16 &
  \cellcolor[HTML]{C8E9E5}8.2\% &
  19 &
  \cellcolor[HTML]{D9F0EE}5.5\% &
  \sayit{Give me the population density for each of these and the standard deviation compared to the mean.} (S6) \\
Filter &
  0 &
  \cellcolor[HTML]{FFFFFF}0.0\% &
  4 &
  \cellcolor[HTML]{F2FAF9}2.1\% &
  4 &
  \cellcolor[HTML]{F7FCFC}1.2\% &
  \sayit{Which counties in this state have higher population density than the average population density of the state?} (S3) \\
Sort &
  3 &
  \cellcolor[HTML]{F1FAF9}2.0\% &
  10 &
  \cellcolor[HTML]{DDF1EF}5.1\% &
  13 &
  \cellcolor[HTML]{E5F5F3}3.8\% &
  \sayit{Tell me the top 5 states lacking broadband access.} (B1) \\
Cluster &
  0 &
  \cellcolor[HTML]{FFFFFF}0.0\% &
  0 &
  \cellcolor[HTML]{FFFFFF}0.0\% &
  0 &
  \cellcolor[HTML]{FFFFFF}0.0\% &
   \\
Pattern &
  28 &
  \cellcolor[HTML]{7CCAC2}18.7\% &
  50 &
  \cellcolor[HTML]{51B8AE}25.6\% &
  78 &
  \cellcolor[HTML]{63C0B7}22.6\% &
  \sayit{Is there a pattern in this region regarding broadband access?} (B1) \\
Outlier &
  0 &
  \cellcolor[HTML]{FFFFFF}0.0\% &
  2 &
  \cellcolor[HTML]{F9FDFC}1.0\% &
  2 &
  \cellcolor[HTML]{FBFEFE}0.6\% &
  \sayit{Can you find significant outliers to population density?} (S5) \\
Legend &
  0 &
  \cellcolor[HTML]{FFFFFF}0.0\% &
  0 &
  \cellcolor[HTML]{FFFFFF}0.0\% &
  0 &
  \cellcolor[HTML]{FFFFFF}0.0\% &
   \\
Shape &
  5 &
  \cellcolor[HTML]{E8F6F5}3.3\% &
  1 &
  \cellcolor[HTML]{FCFEFE}0.5\% &
  6 &
  \cellcolor[HTML]{F3FBFA}1.7\% &
  \sayit{What's the shape of Alabama?} (B6) \\
Spatial Relationships &
  8 &
  \cellcolor[HTML]{DAF0EE}5.3\% &
  0 &
  \cellcolor[HTML]{FFFFFF}0.0\% &
  8 &
  \cellcolor[HTML]{EFF9F8}2.3\% &
  \sayit{What are Florida's neighboring states?} (B5) \\
Visualization Knowledge &
  0 &
  \cellcolor[HTML]{FFFFFF}0.0\% &
  2 &
  \cellcolor[HTML]{F9FDFC}1.0\% &
  2 &
  \cellcolor[HTML]{FBFEFE}0.6\% &
  \sayit{What's a choropleth map?} (S4) \\
General Knowledge &
  38 &
  \cellcolor[HTML]{4DB6AC}25.3\% &
  51 &
  \cellcolor[HTML]{4DB6AC}26.2\% &
  89 &
  \cellcolor[HTML]{4DB6AC}25.8\% &
  \sayit{Why do Midwestern states use predominantly gas?} (B2) \\
Others &
  0 &
  \cellcolor[HTML]{FFFFFF}0.0\% &
  7 &
  \cellcolor[HTML]{E7F5F4}3.6\% &
  7 &
  \cellcolor[HTML]{F1FAF9}2.0\% &
  \sayit{Can you highlight the six states on the map? (S1)} \\ 
\end{tabular}%
\vspace{1em}
\caption{
The distribution of query types by user group, participants (N = 12) asked \sysname 346 questions in total.
Both user groups showed similar usage patterns for the three most common query types (General Knowledge, Pattern, and Action), while BLV users more frequently asked visual queries (Shape, Spatial Relationships) and Retrieve queries compared to sighted users.
This table categorizes all questions based on query type, regardless of whether the system successfully answered them.
}
\label{tab:study-question-types}
\vspace{-2em}
\end{table*}

%% file: tables/table-error-types.tex
\begin{table*}[t]
\centering
\renewcommand{\arraystretch}{1.3}
\footnotesize
\begin{tabular}{@{}l r r 
>{\raggedright\arraybackslash}p{3.5cm}
>{\raggedright\arraybackslash}p{6.5cm}
>{\raggedright\arraybackslash}p{3cm}@{}}
\rowcolor[HTML]{FFFFFF} 
\textbf{Error Type} &
  \multicolumn{1}{l}{\cellcolor[HTML]{FFFFFF}\textbf{Count}} &
  \multicolumn{1}{l}{\cellcolor[HTML]{FFFFFF}\textbf{\%}} &
  \textbf{Example Question} &
  \textbf{\sysname's Incorrect Answer} &
  \textbf{Error Analysis} \\ \midrule
\rowcolor[HTML]{EFF2F6} 
Input Classifier &
  0 &
  0.0\% &
  & & \\
\rowcolor[HTML]{FFFFFF} 
Query Refiner &
  15 &
  26.8\% &
  \sayit{Whats the population density here?} &
  To determine population density, divide the total population of an area by its land area (in square miles or kilometers). For example, if a city has 100,000 people and covers 50 square miles, the population density is 2,000 people per square mile. &
  Failed to catch the focus state {[}`Washington'{]}. \\
\rowcolor[HTML]{EFF2F6} 
Scope Assessor &
  12 &
  21.4\% &
  \sayit{What is the percentage of underserved populations here?} &
  Underserved populations refer to groups lacking adequate access to services, often including low-income, minority, or rural communities. The percentage varies by region and context, but specific statistics can be found in reports from organizations like the U.S. Census Bureau or local health departments. &
  Incorrectly determined this question to be out of scope. \\
\rowcolor[HTML]{FFFFFF} 
Query Processor &
  3 &
  5.4\% &
  \sayit{Why do rural counties use more electricity than gas?} &
  Here's a breakdown of predominantly used heating fuels in Georgia. For urban counties: 75.0\% predominantly use electricity, 25.0\% predominantly use gas. For rural counties: 97.2\% predominantly use electricity, 2.8\% predominantly use gas. &
  Incorrectly classified this question as \textit{compare} when it should have been \textit{general knowledge}. \\
\rowcolor[HTML]{EFF2F6} 
Others &
  26 &
  46.4\% &
  \sayit{Can you create a heat map for the gas heating fuel type layer?} &
  I can't create visual content like heat maps. &
  User asking for additional visualizations that is beyond the system current capabilities. \\
\end{tabular}%
\caption{
Error analysis of incorrectly answered questions by pipeline component. The table shows 56 errors (16.2\% of 346 total questions) across system components, with representative examples of failed queries, incorrect responses, and explanations.
}
\label{tab:error-types}
\vspace{-2em}
\end{table*}

%% file: sections/7-discussion.tex
\section{Discussion \& Future Work}
Below, we discuss key findings, identify challenges, and suggest directions for future work.

\textbf{Capturing and Verbalizing Spatial Patterns.}
Our evaluation revealed substantial overlap in pattern recognition between BLV and sighted participants, demonstrating that our approach represents a step toward creating shared understanding of geospatial patterns.
Nevertheless, we observed important differences highlighting improvement opportunities. 
For instance, S5 incorrectly identified New Mexico as having the ``\textit{highest percentage (of underserved population)}'' when West Virginia actually held this distinction. 
This misinterpretation is likely influenced by New Mexico's large visual area on the map (\autoref{fig:task1}).
BLV users had to perform additional queries about spatial relationships and navigate multiple times around the map to confirm spatial connections, as the pattern summaries only included two states per cluster (a design choice to avoid information overload).
While our implementation relied on LISA clusters and Moran's I~\cite{anselin_local_1995}, all traditional cluster identification methods, (Moran's I~\cite{anselin_local_1995}, Geary's C~\cite{anselin_local_2019}, Getis-Ord Gi*~\cite{getis_analysis_1992}) have limitations in accurately representing patterns that would be visually apparent to sighted users~\cite{wong_issues_2021}.
Future work should explore pattern detection approaches that incorporate estimate error and allow user-defined thresholds, making descriptions more intuitively meaningful and practically valuable~\cite{wong_issues_2021}.

\textbf{Communicating Accuracy.}
Our evaluation revealed that although BLV participants generally trusted the system's answers, they lacked easy methods to verify responses.
Despite prior work~\cite{kim_exploring_2023} emphasizing the importance of communicating uncertainty levels in AI-generated responses, we deliberately chose not to explicitly distinguish which answers were obtained from the local dataset versus those generated by GPT for two reasons: (1) our pipeline classifiers rely on GPT, which itself can make errors in assessment; 
highlighting only those answers that originated entirely from GPT might inadvertently cause users to place excessive trust in other responses; and
(2) many answers combined local raw data with GPT summaries, making it challenging to communicate the varying uncertainty levels between different components in the same response.
A more effective approach would incorporate source links in responses as suggested by B3 and increasingly implemented by commercial AI tools such as Perplexity~\cite{perplexity_ai_perplexity_2022}.  

\textbf{Incorporating Dynamically Guided Prompting.}
Our study revealed meaningful interaction with visualization QA systems requires some level of data literacy.
In several cases, participants struggled not because of system limitations but how they formulated their queries. 
For example, multiple participants asked about \say{fuel} instead of \say{heating fuel} and consequently received explanations about automotive fuel rather than household energy sources.
These challenges highlight the need for dynamic guided prompting to avoid query ambiguity, where the system proactively clarifies user intent during the querying process. 
As S1 observed, \sayit{the system has the potential to correct misinterpretations}. 
Such prompting could be implemented with simple clarifying questions such as: \sayit{Are you asking about fuel consumption in general or specifically about household heating fuel?} 
The system could also consider users' current map focus to offer contextually relevant guided questions.

\textbf{Supporting Direct Manipulation.}
Both screen-reader and sighted participants expressed interest in more direct manipulation capabilities, such as selecting multiple states simultaneously for comparison and analysis.
This desire for additional direct manipulation reflects broader limitations in purely language-based interfaces.
These limitations echo challenges identified decades ago with command-line interfaces: indirect engagement, semantic distance, and articulatory distance~\cite{masson_directgpt_2024}. 
Recent work such as DirectGPT~\cite{masson_directgpt_2024} and Textoshop~\cite{masson_textoshop_2025} has begun exploring ways to integrate direct manipulation with LLM-powered interfaces.
Implementing higher-level direct manipulation capabilities for geovisualization QA systems is an exciting direction for future work and would benefit all user groups.

\textbf{Supporting Users with Diverse Abilities and Expertise.}
Our study revealed that while screen-reader and sighted users have distinct needs, there was considerable overlap in how both groups benefited from \sysname.
We observed that sighted participants demonstrated varying levels of system usage based on their domain expertise and geographic knowledge. 
This finding challenges a binary view of accessibility and suggests that the ability to accurately read, analyze, and interpret geovisualizations applies to a broad spectrum of users. 
When asked to rate the additional benefits of \sysname compared to viewing the visualizations alone, sighted participants gave it a median score of 6 on a scale of 1 (no additional benefit) to 7 (significant improvement). 
This aligns with prior work on designing voice interfaces, which argues that \sayit{when the blind lead the sighted through voice interface design, both blind and sighted users can benefit}~\cite{abdolrahmani_blind_2019}. 
We acknowledge that our screen-reader user group included only blind users and no low-vision users.
Low-vision individuals might rely on hybrid strategies such as combining screen magnification or high-contrast modes with conversational queries.
Future work should investigate how users across this fine-grained spectrum of visual abilities interact with geovisualization QA systems differently, as these insights could inform more nuanced design recommendations for assistive technologies that support diverse perceptual needs and preferences. 
\textbf{Broader Deployment and Scalability.}
While our study focused on U.S. datasets, \sysname's architecture is designed for broader applicability across diverse geospatial domains. 
Currently, the system requires minimal data preprocessing—for all the interactive maps used in the study, researchers simply matched geospatial boundary data with census tables.
Deployment barriers could be further reduced by automating the system's configuration phase.
Specifically, this could involve standardizing a metadata schema to automatically ingest data tables, variable definitions and metric types, as well as automating the manual adjustments currently needed to ensure natural navigation interactions.

\sysname's modular pipeline architecture can be deployed as a browser plugin for existing geovisualization platforms. 
By extracting underlying data and geographic boundaries from web-based maps, the system could augment any choropleth or statistical map with natural language querying, allowing users to retrofit accessibility features without complete interface redesigns.

The system's reliance on general-purpose LLMs for query processing and pattern detection, rather than domain-specific trained models, supports scalability across different contexts through few-shot prompting with domain-appropriate examples. 
Whether applied to urban planning, environmental monitoring, or economic datasets, our findings suggest that this fundamental approach—combining spatial data analysis with natural language interfaces—provides a scalable foundation for making geovisualizations accessible across diverse user populations.

%% file: sections/8-conclusion.tex
\section{Conclusion}
We introduced \sysname, an LLM-based question-answering system that makes geovisualizations accessible to screen-reader users while also benefiting sighted users. Our evaluation with twelve participants demonstrated that both user groups successfully engaged with geospatial data through different but complementary interaction strategies—screen-reader users combining verbal queries with keyboard navigation, and sighted users leveraging visual assessment with targeted queries. Despite these different approaches, both groups identified similar patterns, showing that our system creates a shared understanding of geovisualizations. \sysname demonstrates that inclusive design enhances spatial data analysis for all users, regardless of visual ability.

%% file: sections/9-acks.tex
\begin{acks}
We are grateful to our co-design collaborators and study participants, without whom this work would not have been possible. 
We also thank Lotus Zhang for assistance with participant recruitment, Maurício Sousa for valuable feedback, and Sandy Kaplan for guidance on academic writing. 
This work was supported by NSF SCC-IRG \#2125087.
\end{acks}